\documentstyle{article}

\catcode`\@=11
\def\@citex[#1]#2{%
\if@filesw \immediate \write \@auxout {\string \citation {#2}}\fi
\@tempcntb\m@ne \let\@h@ld\relax \def\@citea{}%
\@cite{%
  \@for \@citeb:=#2\do {%
    \@ifundefined {b@\@citeb}%
      {\@h@ld\@citea\@tempcntb\m@ne{\bf ?}%
      \@warning {Citation `\@citeb ' on page \thepage \space undefined}}%
      {\@tempcnta\@tempcntb \advance\@tempcnta\@ne%
      \@tempcntb\number\csname b@\@citeb \endcsname \relax%
      \ifnum\@tempcnta=\@tempcntb 
        \ifx\@h@ld\relax%
          \edef \@h@ld{\@citea\csname b@\@citeb\endcsname}%
        \else%
          \edef\@h@ld{\ifmmode{-}\else--\fi\csname b@\@citeb\endcsname}%
        \fi%
      \else
        \@h@ld\@citea\csname b@\@citeb \endcsname%
        \let\@h@ld\relax%
      \fi}%
    \def\@citea{,\penalty\@highpenalty\,}%
  }\@h@ld
}{#1}}

\def\@citeb#1#2{{[#1]\if@tempswa , #2\fi}}
\def\@citeu#1#2{{$^{#1}$\if@tempswa , #2\fi }}
\def\@citep#1#2{{#1\if@tempswa , #2\fi}}

\def\bcites{         
        \catcode`\@=11
        \let\@cite=\@citeb
        \catcode`\@=12
}

\def\upcites{         
        \catcode`\@=11
        \let\@cite=\@citeu
        \catcode`\@=12
}

\def\plaincites{      
        \catcode`\@=11
        \let\@cite=\@citep
        \catcode`\@=12
}

\newcount\hour
\newcount\minute
\newtoks\amorpm
\hour=\time\divide\hour by 60
\minute=\time{\multiply\hour by 60 \global\advance\minute by-\hour}
\edef\standardtime{{\ifnum\hour<12 \global\amorpm={am}%
        \else\global\amorpm={pm}\advance\hour by-12 \fi
        \ifnum\hour=0 \hour=12 \fi
        \number\hour:\ifnum\minute<10 0\fi\number\minute\the\amorpm}}
\edef\militarytime{\number\hour:\ifnum\minute<10 0\fi\number\minute}

\def\draftlabel#1{{\@bsphack\if@filesw {\let\thepage\relax
   \xdef\@gtempa{\write\@auxout{\string
      \newlabel{#1}{{\@currentlabel}{\thepage}}}}}\@gtempa
   \if@nobreak \ifvmode\nobreak\fi\fi\fi\@esphack}
        \gdef\@eqnlabel{#1}}
\def\@eqnlabel{}
\def\@vacuum{}
\def\marginnote#1{}
\def\draftmarginnote#1{\marginpar{\raggedright\scriptsize\tt#1}}
\def\draft{
        \pagestyle{plain}
        \overfullrule=2pt
        \oddsidemargin -.5truein
        \def\@oddhead{\sl \phantom{\today\quad\militarytime} \hfil
        \smash{\Large\sl DRAFT} \hfil \today\quad\militarytime}
        \let\@evenhead\@oddhead
        \let\label=\draftlabel
        \let\marginnote=\draftmarginnote
        \def\ps@empty{\let\@mkboth\@gobbletwo
        \def\@oddfoot{\hfil \smash{\Large\sl DRAFT} \hfil}
        \let\@evenfoot\@oddhead}
        \def\@eqnnum{(\theequation)\rlap{\kern\marginparsep\tt\@eqnlabel}%
        \global\let\@eqnlabel\@vacuum}  }

\def\blackfonts{
        \font\blackboard=msbm10 scaled\magstep1
        \font\blackboards=msbm8
        \font\blackboardss=msbm6
}

\def\prep{         
        \catcode`\@=11
        \input art10.sty
        \catcode`\@=12
        
        \let\small\null
        \def\blackfonts{
                \font\blackboard=msbm10
                \font\blackboards=msbm7
                \font\blackboardss=msbm5
        }
        \let\sl\it
        \twocolumn
        \sloppy
        \voffset=-2.54truecm
        \hoffset=-2.54truecm
        \flushbottom
        \parindent 1em
        \leftmargini 2em
        \leftmarginv .5em
        \leftmarginvi .5em
        \marginparwidth 48pt
        \marginparsep 10pt
        \setlength{\columnsep}{2truecm}
        \setlength{\textwidth}{25.4truecm}
        \setlength{\textheight}{17truecm}
        \baselineskip=16pt
        \oddsidemargin .18truein
        \evensidemargin .17truein
}

\def\eqalign#1{\null\,\vcenter{\openup\jot\m@th
  \ialign{\strut\hfil$\displaystyle{##}$&$\displaystyle{{}##}$\hfil
      \crcr#1\crcr}}\,}
\def\eqalignno#1{\displ@y \tabskip\centering
  \halign to\displaywidth{\hfil$\@lign\displaystyle{##}$\tabskip\z@skip
    &$\@lign\displaystyle{{}##}$\hfil\tabskip\centering
    &\llap{$\@lign##$}\tabskip\z@skip\crcr
    #1\crcr}}

\def\section{\@startsection {section}{1}{\z@}{3.ex plus 1ex minus
 .2ex}{2.ex plus .2ex}{\large\bf}}
\def\subsection{\@startsection{subsection}{2}{\z@}{2.75ex plus 1ex minus
 .2ex}{1.5ex plus .2ex}{\bf}}        

\def\appendix{{\newpage\section*{Appendix}}\let\appendix\section%
        {\setcounter{section}{0}
        \gdef\thesection{\Alph{section}}}\section}

\def\abstract{\if@twocolumn
\section*{Abstract}
\else 
\begin{center}
{\bf Abstract\vspace{-.5em}\vspace{0pt}}
\end{center}
\quotation
\fi}

\catcode`\@=12

\def\d{\partial}

\def\sqr#1#2{{\vcenter{\vbox{\hrule height.#2pt\hbox{\vrule width.#2pt 
height#1pt \kern#1pt \vrule width.#2pt}\hrule height.#2pt}}}}

\def\=d{\,{\buildrel\rm def\over =}\,}

\def\F{{\cal F}}

\def\i3p{\p32\int d^3p}

\def\As{A\hbox to 1pt{\hss /}}
\def\np4{\int d^4p_1\cdots d^4p_{n-1}\, }

\def\Tr{{\rm Tr}\, }

\def\nx4{\int d^4x_1\ldots d^4x_n\, }

\def\kon#1#2{\vbox{\halign{##&&##\cr
\lower4pt\hbox{$\scriptscriptstyle\vert$}\hrulefill &
\hrulefill\lower4pt\hbox{$\scriptscriptstyle\vert$}\cr $#1$&
$#2$\cr}}}

\def\konv#1#2#3{\hbox{\vrule height12pt depth-1pt}
\vbox{\hrule height12pt width#1cm depth-11.6pt}
\hbox{\vrule height6.5pt depth-0.5pt}
\vbox{\hrule height11pt width#2cm depth-10.6pt\kern5pt
      \hrule height6.5pt width#2cm depth-6.1pt}
\hbox{\vrule height12pt depth-1pt}
\vbox{\hrule height6.5pt width#3cm depth-6.1pt}
\hbox{\vrule height6.5pt depth-0.5pt}}
\def\konu#1#2#3{\hbox{\vrule height12pt depth-1pt}
\vbox{\hrule height1pt width#1cm depth-0.6pt}
\hbox{\vrule height12pt depth-6.5pt}
\vbox{\hrule height6pt width#2cm depth-5.6pt\kern5pt
      \hrule height1pt width#2cm depth-0.6pt}
\hbox{\vrule height12pt depth-6.5pt}
\vbox{\hrule height1pt width#3cm depth-0.6pt}
\hbox{\vrule height12pt depth-1pt}}

\def\konw#1#2#3{\hbox{\vrule height12pt depth-1pt}
\vbox{\hrule height12pt width#1cm depth-11.6pt}
\hbox{\vrule height6.5pt depth-0.5pt}
\vbox{\hrule height12pt width#2cm depth-11.6pt \kern5pt
      \hrule height6.5pt width#2cm depth-6.1pt}
\hbox{\vrule height6.5pt depth-0.5pt}
\vbox{\hrule height12pt width#3cm depth-11.6pt}
\hbox{\vrule height12pt depth-1pt}}

\def\i{{\rm int}}

\def\c{{\rm cl}}
\def\e{{\rm ext}}

\def\r{{\rm ret}}

\def\m3{{\mu_1\mu_2\mu_3}}

\def\co{{\rm Com}}

\def\p{{(+)}}

\def\be{\begin{equation}}       \def\eq{\begin{equation}}
\def\ee{\end{equation}}         \def\eqe{\end{equation}}

\def\bea{\begin{eqnarray}}      \def\eqa{\begin{eqnarray}}
\def\ena{\end{eqnarray}}        \def\eea{\end{eqnarray}}
                                \def\eqae{\end{eqnarray}}

\def\ba{\begin{array}}
\def\ea{\end{array}}
\def\unit{1 \hskip-.3em \raise2pt\hbox{$ \scriptstyle |$ } }

\def\c{\gamma} 
\def\d{\delta}
\def\e{\epsilon}           
\def\f{\phi}               
\def\g{\gamma}
   
\def\i{\iota}

\def\k{\kappa}                    
\def\l{\lambda}
\def\m{\mu}
\def\n{\nu}
  
\def\p{\pi}                
\def\r{\rho}                                     
\def\s{\sigma}                                   
\def\t{\tau}

\def\D{\Delta}
\def\F{\Phi}
\def\G{\Gamma}

\def\L{\Lambda}

\def\ca{{\cal A}}

\def\co{{\cal O}}

\def\half{{1 \over 2}}

\def\bop#1{\setbox0=\hbox{$#1M$}\mkern1.5mu
        \vbox{\hrule height0pt depth.04\ht0
        \hbox{\vrule width.04\ht0 height.9\ht0 \kern.9\ht0
        \vrule width.04\ht0}\hrule height.04\ht0}\mkern1.5mu}
\def\Box{{\mathpalette\bop{}}}                        
\def\pa{\partial}                              

\def\>{\rangle} 

\def\<{\langle} 
\def\Dsl{D \hskip-.6em \raise1pt\hbox{$ / $ } }

\def\sl#1{\rlap{\hbox{$\mskip 1 mu /$}}#1}
\def\leftrightarrowfill{$\mathsurround=0pt \mathord\leftarrow \mkern-6mu
       \cleaders\hbox{$\mkern-2mu \mathord- \mkern-2mu$}\hfill
       \mkern-6mu \mathord\rightarrow$}
\def\dvec#1{\vbox{\ialign{##\crcr
       \leftrightarrowfill\crcr\noalign{\kern-1pt\nointerlineskip}
       $\hfil\displaystyle{#1}\hfil$\crcr}}}          
\def\hook#1{{\vrule height#1pt width0.4pt depth0pt}}
\def\leftrighthookfill#1{$\mathsurround=0pt \mathord\hook#1
       \hrulefill\mathord\hook#1$}
\def\underhook#1{\vtop{\ialign{##\crcr                 
       $\hfil\displaystyle{#1}\hfil$\crcr
       \noalign{\kern-1pt\nointerlineskip\vskip2pt}
       \leftrighthookfill5\crcr}}}
\def\smallunderhook#1{\vtop{\ialign{##\crcr      
       $\hfil\scriptstyle{#1}\hfil$\crcr
       \noalign{\kern-1pt\nointerlineskip\vskip2pt}
       \leftrighthookfill3\crcr}}}

\def\sfrac#1#2{{\vphantom1\smash{\lower.5ex\hbox{\small$#1$}}\over
       \vphantom1\smash{\raise.4ex\hbox{\small$#2$}}}} 
\def\bfrac#1#2{{\vphantom1\smash{\lower.5ex\hbox{$#1$}}\over
       \vphantom1\smash{\raise.3ex\hbox{$#2$}}}}      
\def\afrac#1#2{{\vphantom1\smash{\lower.5ex\hbox{$#1$}}\over#2}}  
\def\on#1#2{{\buildrel{\mkern2.5mu#1\mkern-2.5mu}\over{#2}}}
\def\ddt#1{\on{\hbox{\LARGE .\kern-2pt.}}#1}             
\def\tdt#1{\on{\hbox{\LARGE .\kern-2pt.\kern-2pt.}}#1}   

\def\boxes#1{
       \newcount\num
       \num=1
       \newdimen\downsy
       \downsy=-1.5ex
       \mskip-2.8mu
       \bo
       \loop
       \ifnum\num<#1
       \llap{\raise\num\downsy\hbox{$\bo$}}
       \advance\num by1
       \repeat}
\def\boxup#1#2{\newcount\numup
       \numup=#1
       \advance\numup by-1
       \newdimen\upsy
       \upsy=.75ex
       \mskip2.8mu
       \raise\numup\upsy\hbox{$#2$}}

\newskip\humongous \humongous=0pt plus 1000pt minus 1000pt
\def\caja{\mathsurround=0pt}
\def\eqalign#1{\,\vcenter{\openup2\jot \caja
       \ialign{\strut \hfil$\displaystyle{##}$&$
       \displaystyle{{}##}$\hfil\crcr#1\crcr}}\,}
\newif\ifdtup

\def\to{\rightarrow}

\def\1ov4{{1\over 4}}

\def\Tr{{\rm Tr}}

\def\pa{\partial}

\def\ddt{\dot{\t}}

\def\pa{\partial}

\def\nonu{\nonumber \\{}}
\def\half{{1 \over 2}}

\def\g0{g_{(0)}}
\def\gi{g_{(0)}^{-1}}
\newcommand{\sm}[1]{\mbox{\scriptsize #1}} 
\newcommand{\tnnn}[1]{\mbox{\tiny #1}} 
\def\GN{G_{\mbox{\tnnn N}}}
\def\d{\mbox{d}} 
\renewcommand{\theequation}{\thesection.\arabic{equation}}

\begin{document}

\null\vskip-24pt
\begin{flushright}
SPIN-2000/05\\
ITP-UU-00/03\\
PUTP-1921 \\
{\tt hep-th/0002230}
\end{flushright}
\vskip0.3truecm
\begin{center}
\vskip 1truecm
{\Large\bf
Holographic Reconstruction of Spacetime}  \\
\vspace{.3cm}
{\Large\bf and Renormalization in the AdS/CFT Correspondence}\\ 
\vskip 1.5truecm

{\large\bf Sebastian de Haro,${}^{\star\dagger}${}\footnote{
e-mail: {\tt haro@phys.uu.nl}}
Kostas Skenderis${}^{\ddagger}$\footnote{
e-mail: {\tt kostas@feynman.princeton.edu}}
and Sergey N. Solodukhin${}^{\star}${}\footnote{
e-mail: {\tt S.Solodukhin@phys.uu.nl}}}
\\
\vskip 1truecm
${}^{\star}$ {\it Spinoza Institute, 
Utrecht University\\
Leuvenlaan 4, 3584 CE Utrecht, The Netherlands}
\vskip 1truemm
${}^{\dagger}$ {\it Institute for Theoretical Physics, 
Utrecht University\\
Princetonplein 5, 3584 CC
Utrecht, The Netherlands}
\vskip 1truemm
${}^{\ddagger}$ {\it Physics Department, 
Princeton University \\
Princeton, NJ 08544, USA}

\end{center}
\vskip .5truecm
\begin{center}
{\bf\large Abstract}
\end{center}
\normalsize
We develop a systematic method for renormalizing 
the AdS/CFT prescription for computing correlation functions.
This involves regularizing the bulk on-shell supergravity 
action in a covariant way, computing all divergences,
adding counterterms to cancel them and then removing 
the regulator. We explicitly work out the case of pure 
gravity up to six dimensions and of gravity coupled to 
scalars. The method can also be viewed as providing
a holographic reconstruction of the bulk spacetime metric and of 
bulk fields on this spacetime, out of conformal field theory data.
Knowing which sources are turned on 
is sufficient in order to obtain an asymptotic expansion 
of the bulk metric and of bulk fields near the boundary to high enough 
order so that all infrared divergences of the 
on-shell action are obtained. To continue the holographic
reconstruction of the bulk fields one needs
new CFT data: the expectation value of the 
dual operator. In particular, in order to obtain the bulk 
metric one needs to know the expectation value of  
stress-energy tensor of the boundary theory.
We provide completely explicit formulae
for the holographic stress-energy tensors up to six dimensions.
We show that both the gravitational and matter conformal 
anomalies of the boundary theory are correctly reproduced.
We also obtain the conformal transformation
properties of the boundary stress-energy tensors.

\section{Introduction and summary of the results}
\setcounter{equation}{0}

Holography states that a $(d{+}1)$-dimensional gravitational 
theory (referred to as the bulk theory) should have a 
description in terms of a $d$-dimensional
field theory (referred to as the boundary theory)
with one degree of freedom per Planck area
\cite{tHooft,Susskind}. The arguments leading to the 
holographic principle use rather generic properties
of gravitational physics, indicating that holography
should be a feature of any quantum theory of gravity.
Nevertheless it has been proved a difficult task 
to find examples where holography is realized,
let alone to develop a precise dictionary 
between bulk and boundary physics. The AdS/CFT 
correspondence \cite{Malda} provides such a realization \cite{Wit,SussWit}
with a rather precise computational framework \cite{Gubs,Wit}.
It is, therefore, desirable to sharpen the existing 
dictionary between bulk/boundary physics as much as possible. 
In particular, one of the issues one would like to
understand is how spacetime is built 
holographically out of field theory data.

The prescription of \cite{Gubs,Wit} gives a concrete 
proposal for a holographic computation of physical
observables. In particular, the partition function
of string theory compactified on AdS spaces 
with prescribed boundary conditions for the 
bulk fields is equal to the generating functional 
of conformal field theory correlation 
functions, the boundary value of fields being now 
interpreted as sources for operators of the dual conformal 
field theory (CFT). 
String theory on anti-de Sitter (AdS) spaces is still incompletely
understood. At low energies, however, the theory 
becomes a gauged supergravity with an AdS ground
state coupled to Kaluza-Klein (KK) modes. On the 
field theory side, this corresponds to the large $N$
and strong 't Hooft coupling regime of the 
CFT. So in the AdS/CFT context the question is how
one can reconstruct the bulk spacetime out of
CFT data. One can also pose the converse 
question: given a bulk spacetime, what 
properties of the dual CFT can one read off?

The prescription of \cite{Gubs,Wit} equates the 
on-shell value of the supergravity action 
with the generating functional of connected graphs
of composite operators. Both sides of this 
correspondence, however, suffer from infinities
---infrared divergences on the supergravity side
and ultraviolet divergences on the CFT side.
Thus, the prescription of \cite{Gubs,Wit}
should more properly be viewed as an equality between
bare quantities. Ones needs to renormalize the theory to obtain a 
correspondence between 
finite quantities. It is one of the aims of this 
paper to present a systematic way of performing 
such renormalization.
 
The CFT data\footnote{We assume that the 
CFT we are discussing has an AdS dual.
Our results only depend on the spacetime 
dimension and apply to all cases where the AdS/CFT duality 
is applicable, so we shall not specify any particular
CFT model.} that we will use are: which 
operators are turned on, and what is their vacuum 
expectation value. Since the boundary metric (or, more properly,
the boundary conformal structure) couples to the boundary 
stress-energy tensor, the reconstruction of the 
bulk metric to leading order involves a detailed
knowledge of the way the energy-momentum  tensor 
is encoded holographically. 
There is by now an extended literature on the 
study of the stress-energy tensor in the context of the 
AdS/CFT correspondence starting from \cite{BK,Myers}.
We will build on these and other related works \cite{EJM,Mann,KLS}. 
Our starting point will be the calculation of the 
infrared divergences of the on-shell gravitational action
\cite{HS}. Minimally subtracting the divergences 
by adding counterterms \cite{HS} leads straightforwardly to the 
results in \cite{BK,EJM,KLS}. After the subtractions 
have been made one can remove the (infrared) regulator
and obtain a completely explicit formula for
the expectation value of the dual 
stress-energy tensor in terms of the gravitational solution.

We will mostly concentrate on the gravitational sector,
i.e. in the reconstruction of the bulk metric, 
but we will also discuss the coupling to scalars. 
Our approach will be to build perturbatively an Einstein manifold 
of constant negative curvature (which we will sometimes 
refer to as an asymptotically 
AdS space) as well as a solution to the scalar field equations
on this manifold out of CFT data. The CFT data we start from 
is what sources are turned on. We will include 
a source for the dual stress-energy tensor as well 
as sources for scalar composite operators. 
This means that in the bulk we need to solve the 
gravitational equations coupled to scalars 
given a conformal structure at infinity and 
appropriate Dirichlet boundary conditions for the 
scalars. It is well-known that if one considers the standard Euclidean AdS 
(i.e., with isometry $SO(1,d+1)$), the scalar field
equation with Dirichlet boundary conditions 
has a unique solution. In the Lorentzian case,
because of the existence of normalizable modes, the 
solution ceases to be unique. Likewise,
the Dirichlet boundary condition problem 
for (Euclidean) gravity has a unique (up to diffeomorphisms)
smooth solution in the case the bulk manifold in 
topologically a ball and the boundary conformal 
structure sufficiently close to the standard one \cite{GrahamLee}.
However, given a boundary topology there may be 
more than one Einstein manifold with this boundary. 
For example, if the boundary has the topology
of $S^1 \times S^{d-1}$, there are two possible bulk manifolds
\cite{PageH,Wit}:
one which is obtained from standard AdS by global identifications
and is topologically $S^1 \times R^d$,
and another, the Schwarzschild-AdS black hole, 
which is topologically $R^2 \times S^{d-1}$.

We will make no assumption
on the global structure of the space or on its
signature. The CFT should provide additional 
data in order to retrieve this information.
Indeed, we will see that only the information 
about the sources leaves undetermined the 
part of the solution which is sensitive on 
global issues and/or the signature of spacetime.
To determine that part one needs new CFT data.
To leading order these are  
the expectation values of the CFT operators.

In particular, in the case of pure gravity, we find that 
generically a boundary conformal structure 
is not sufficient in order to 
obtain the bulk metric. One needs more CFT data.
To leading order one needs to specify  
the expectation value of the boundary stress-energy tensor.
Since the gravitational field equation is a second
order differential equation, one may expect that these data are sufficient 
in order to specify the full solution.
In general, however, non-local observables such as Wilson loops
may be needed in order to recover global properties of the solution and
reconstruct the metric in the deep interior region. Furthermore, 
higher point functions of the stress-energy tensor
may be necessary if higher derivatives corrections
such as $R^2$ terms are included in the action.
We emphasize that we make no assumption about the regularity 
of the solution. Under additional assumptions 
the metric may be determined by fewer data. 
For example, as we mentioned above, under certain 
assumptions on the topology and the 
boundary conformal structure one obtains 
a unique smooth solution \cite{GrahamLee}.
Another example is the case when
one restricts oneself to conformally
flat bulk metrics. Then a conformally flat boundary metric 
does yield a unique, up to diffeomorphisms and 
global identifications, bulk metric \cite{SkSo}.

Turning things around, given a specific solution,
we present formulae for the expectation values 
of the dual CFT operators. In particular, in the 
case the operator is the stress energy tensor,
our formulae have a ``dual'' meaning \cite{BK}:
both as the expectation value of the 
stress-energy tensor of the dual CFT 
and as the quasi-local stress-energy tensor 
of Brown and York \cite{BrownYork}. We provide very explicit
formulae for the stress-energy tensor associated with
any solution of Einstein's equations with 
negative constant curvature. 

Let us summarize these results for spacetime 
dimension up to six. The first step is to 
rewrite the solution in the Graham-Fefferman
coordinate system \cite{FeffermanGraham}
\be \label{GrFe}
ds^2=G_{\m \n} dx^\m dx^\n = {l^2 \over r^2}\left(dr^2 + 
g_{ij}(x,r) dx^i dx^j\right)
\ee
where 
\be
g(x,r)=g_{(0)} + r^2 g_{(2)}+ \cdots + r^d g_{(d)} + h_{(d)} r^{d} \log r^2 +
\co(r^{d+1})
\ee
The logarithmic term appears only in even dimensions and
only even powers of $r$ appear up to order $r^{[(d-1)]}$, where 
$[a]$ indicates the integer part of $a$. $l$ is a parameter of 
dimension of length related to the cosmological constant as 
$\L=-{d(d-1) \over 2 l^2}$. Any asymptotically AdS metric can be brought in 
the form (\ref{GrFe}) near the boundary (\cite{GrahamLee}, 
see also \cite{GrahamWitten,Graham}). Once this coordinate 
system has been reached, the stress-energy tensor 
reads
\be \label{tx}
\<T_{ij}\>={d l^{d-1} \over 16 \p \GN}\, g_{(d)ij} + X_{ij}[g_{(n)}].
\ee
where $X_{ij}[g_{(n)}]$ is a function of $g_{(n)}$ with $n<d$.
Its exact form depends on the spacetime dimension and it reflects the
conformal anomalies of the boundary conformal field theory.
In odd (boundary) dimensions, where there are no gravitational conformal
anomalies, $X_{ij}$ is equal to zero. The expression for $X_{ij}[g_{(n)}]$
for $d=2,4,6$ can be read off from (\ref{T2}), (\ref{T4}) and (\ref{T6}),
respectively.
The universal part of (\ref{tx}) (i.e. with $X_{ij}$ omitted)  
was obtained previously in \cite{Myers}. 
Actually, to obtain the dual stress-energy tensor it is 
sufficient to only know $g_{(0)}$ and $g_{(d)}$ as $g_{(n)}$ with 
$n<d$ are uniquely determined from $g_{(0)}$, as we will see. 
The coefficient $h_{(d)}$ of the logarithmic term 
in the case of even $d$ is also directly related to the 
conformal anomaly: it is proportional to the metric 
variation of the conformal anomaly.
 
It was pointed out in \cite{BK} that this prescription for 
calculating the boundary stress-energy tensor  
provides also a novel, free of divergences\footnote{
We emphasize, however, that one 
has to subtract the logarithmic divergences in even dimensions
in order for the stress-energy tensor to be finite.},
way of computing the gravitational quasi-local 
stress-energy tensor of Brown and York \cite{BrownYork}.
This approach was recently criticized in \cite{AshDas},
and we take this opportunity to address this criticism.
Conformal anomalies reflect infrared divergences 
in the gravitational sector \cite{HS}. 
Because of these divergences
one cannot maintain the full group of isometries even asymptotically.
In particular, the isometries of AdS that rescale the radial 
coordinate (these correspond to dilations in the CFT) 
are broken by infrared divergences. 
Because of this fact, 
bulk solutions that are related by diffeomorphisms that
yield a conformal transformation in the boundary do not 
necessarily have the same mass. Assigning zero mass 
to the spacetime with boundary $R^d$,
one obtains that, due to the conformal anomaly,
the solution with boundary $R \times S^{d-1}$ has 
non-zero mass. This parallels exactly the discussion 
in field theory. In that case, starting from the 
CFT on $R^d$ with vanishing expectation value
of the stress-energy tensor, one obtains 
the Casimir energy of the CFT on $R \times S^{d-1}$
by a conformal transformation \cite{CC}. The
agreement between the gravitational ground state energy 
and the Casimir energy of the CFT is a direct consequence
of the fact that the conformal anomaly 
computed by weakly coupled gauge theory and by 
supergravity agree \cite{HS}. It should be noted that, 
as emphasized in \cite{BK}, agreement between gravity/field
theory for the ground state energy is achieved only after
all ambiguities are fixed in the same manner on both
sides.
 
A conformal transformation in the boundary theory is realized in the bulk as a 
special diffeomorphism that preserves the form
of the coordinate system (\ref{GrFe}) \cite{ISTY}.
Using these diffeomorphisms one can easily study 
how the (quantum, i.e., with the effects of the 
conformal anomaly taken into account) stress-energy 
tensor transforms under conformal transformations.
Our results, when restricted to the
cases studied in the literature \cite{CC},
are in agreement with them. We note that the present 
determination is considerably easier than the one in \cite{CC}.

The discussion is qualitatively the same when one 
adds matter to the system. We discuss 
scalar fields but the discussion generalizes straightforwardly
to other kinds of matter. We study both the case the 
gravitational background is fixed and the case
gravity is dynamical. 

Let us summarize the results for the case of scalar fields
in a fixed gravitational background (given by a 
metric of the form (\ref{GrFe})). We look for 
solutions of massive scalar fields with mass
$m^2=(\D-d) \D$
that near the boundary have the form (in the coordinate system (\ref{GrFe}))
\be \label{frexp}
\F(x,r)=r^{d-\D}\left(\f_{(0)} + r^2 \f_{(2)} + \cdots +
r^{2\D-d} \f_{(2\D-d)} + r^{2\D-d} \log r^2 \psi_{(2\D-d)}\right)
+\co(r^{\D+1}).
\ee
The logarithmic terms appears only when $2\D-d$ is an integer
and we only consider this case in this paper.
We find that $\f_{(n)}$, with $n<2\D-d$, and $\psi_{(2\D-d)}$
are uniquely determined from the scalar field equation.
This information is sufficient for a complete 
determination of the infrared divergences of the 
on-shell bulk action. In particular, 
the logarithmic term $\psi_{(2\D-d)}$ in (\ref{frexp})
is directly related to matter conformal anomalies.
These conformal anomalies were shown not to renormalize
in \cite{PeSk}. We indeed find exact agreement with the computation 
in \cite{PeSk}. Adding counterterms to cancel 
the infrared divergences we obtain the renormalized on-shell
action. We stress that even in the case of a free 
massive scalar field in a fixed AdS background 
one needs counterterms in order for the on-shell action
to be finite (see (\ref{finiteact})).
The coefficient $\f_{(2\D-d)}$ is left undetermined
by the field equations. It is determined, however, by the 
expectation value of the dual operator. Differentiating the 
renormalized on-shell action one finds (up to terms 
contributing contact terms in the 2-point function)
\be
\< O (x)\> = (2 \D- d) \f_{(2\D-d)}(x)
\ee
This relation, with the precise proportionality coefficient,
has first been derived in \cite{KleWit}. The value of the proportionality 
coefficient is crucial in order to obtain the correct 
normalization  of the 2-point function in standard
AdS background \cite{FMMR}.

In the case the gravitational background is dynamical we find that, 
for scalars that correspond to irrelevant operators, our
perturbative treatment is consistent only if one considers
single insertions of the irrelevant operator, i.e. the source
is treated as an infinitesimal parameter, in agreement with the discussion 
in \cite{Wit}. For scalars that correspond to marginal and
relevant operators one can compute perturbatively the back-reaction
of the scalars to the gravitational background. One can then 
regularize and renormalize as in the discussion of pure 
gravity or scalars in a fixed background. For illustrative
purposes we analyze a simple example.

This paper is organized as follows. In the next section we discuss 
the Dirichlet problem for AdS gravity and we obtain an asymptotic 
solution for a given boundary metric (up to six dimensions).
In section 3 we use these solutions to obtain 
the infrared divergences of the on-shell gravitational action.
After renormalizing the on-shell action by adding counterterms,
we compute the holographic stress-energy tensor. Section 4 is devoted
to the study of the conformal transformation properties of the 
boundary stress-energy tensor. In section 5 we extend the analysis
of sections 2 and 3 to include matter. 
In appendices \ref{EinSol} and \ref{as-sc}
we give  the explicit form of the solutions discussed 
in section 2 and section 5. Appendix \ref{div-ind}
contains the explicit form of the counterterms discussed 
in section 3. 
Finally, in appendix \ref{h-a} we present a proof that the coefficient
of the logarithmic term in the metric (present in even 
boundary dimensions)
is proportional to the metric variation of the conformal anomaly.

\section{Dirichlet boundary problem for AdS gravity}
\setcounter{equation}{0}

The Einstein-Hilbert action for a theory on a manifold $M$ 
with boundary $\pa M$ is given by\footnote{
Our curvature conventions are as follows
$R_{ijk}{}^l=\pa_i \G_{jk}{}^l + \G_{ip}{}^l \G_{jk}{}^p - i
\leftrightarrow j$ and $R_{ij}=R_{ikj}{}^k$. We these conventions
the curvature of AdS comes out positive, but we will 
still use the terminology ``space of constant negative
curvature''. Notice also that we take
$\int \d^{d+1} x = \int \d^d x \int_0^\infty \d r$
and the boundary is at $r=0$ (in the coordinate system (\ref{GrFe})). 
The minus sign in front of the trace of the second 
fundamental form is correlated with the choice of having $r=0$ in 
the lower end of the radial integration.} 
\be \label{action}
S_{\sm{gr}}[G]={1 \over 16 \p \GN}[\int_{M}\d^{d+1}x\, 
\sqrt{G}\, (R[G] + 2 \L) 
- \int_{\pa M} \d^d x\, \sqrt{\c}\, 2 K],
\ee
where $K$ is the trace of the second fundamental form and
$\c$ is the induced metric on the boundary. The 
boundary term is necessary in order to get an action which 
only depends on first derivatives of the metric \cite{GibbonsHawking},
and it guarantees that the variational 
problem with Dirichlet boundary conditions is well-defined.

According to the prescription of \cite{Gubs,Wit}, the conformal 
field theory effective action is given by evaluating the 
on-shell action functional. The field specifying the 
boundary conditions for the metric is regarded as a source 
for the boundary operator. We therefore need to obtain solutions
to Einstein's equations,
\be \label{feq1}
R_{\m \n} - \half R G_{\m \n} = \L G_{\m \n},
\ee
subject to appropriate Dirichlet boundary conditions.

Metrics $G_{\m \n}$ that satisfy (\ref{feq1}) have a second order 
pole at infinity. Therefore, they do not induce a 
metric at infinity. They do induce, however, a conformal class, i.e. a 
metric up to a conformal transformation. This is achieved by 
introducing a defining function $r$, i.e. a positive function in the 
interior of $M$ that  has a single zero and non-vanishing derivative 
at the boundary. Then one obtains
a metric at the boundary by $g_{(0)}= r^2 G|_{\pa M}$ 
{}\footnote{Throughout this article the metric $g_{(0)}$
is assumed to be non-degenerate. For studies of the AdS/CFT 
correspondence in cases where $g_{(0)}$ is degenerate 
we refer to \cite{BPSV,marika}.}. 
However, any other defining function $r'=r \exp w$ 
is as good. Therefore, the metric $g_{(0)}$ is only defined up to 
a conformal transformation. 

We are interested in solving (\ref{feq1}) given a conformal structure
at infinity. This can be achieved by working in the coordinate 
system (\ref{GrFe}) introduced by Feffermam and Graham \cite{FeffermanGraham}.
The metric in (\ref{GrFe}) 
contains only even powers of $r$ up to the order we are interested in
\cite{FeffermanGraham} (see also \cite{GrahamWitten,Graham}).
For this reason, it is convenient to use the variable $\r=r^2$ \cite{HS},
{}\footnote{Greek indices, $\m,\n,..$ are used for $d+1$-dimensional indices, 
Latin ones, $i,j,..$ for $d$-dimensional ones. 
To distinguish the curvatures of the various metrics introduced in 
(\ref{coord}) we will often use the notation $R_{ij}[g]$ to 
indicate that this is the Ricci tensor of the metric $g$, etc.} 
\bea \label{coord}
&&ds^2=G_{\m \n} dx^\m dx^\n = l^2 \left({d\r^2 \over 4 \r^2} + 
{1 \over \r} g_{ij}(x,\r) dx^i dx^j \right) \nonu
&&g(x,\r)=g_{(0)} + \cdots + \r^{d/2} g_{(d)} + h_{(d)} \r^{d/2} \log \r + ... 
\eea
where the logarithmic piece appears only for even $d$.
The sub-index in the metric expansion (and in all other 
expansions that appear in this paper) indicates the number
of derivatives involved in that term, i.e. $g_{(2)}$ contains 
two derivatives, $g_{(4)}$ four derivatives, etc. It follows
that the perturbative expansion in $\rho$ is also 
a low energy expansion. We set $l=1$ from now on. One can easily 
reinstate the factors of $l$ by dimensional analysis.

One can check that the curvature of $G$ satisfies
\be \label{ads}
R_{\k \l \m \n}[G] = (G_{\k \m} G_{\l \n} - G_{\k \n} G_{\l \m}) + \co(\r)
\ee
In this sense the metric is asymptotically anti-de Sitter. The 
Dirichlet problem for Einstein metrics satisfying (\ref{ads}) 
exactly (i.e. not only to leading order in $\rho$) was solved in \cite{SkSo}.
 
In the coordinate system (\ref{coord}), Einstein's equations read \cite{HS}
\bea
\rho \,[2 g^{\prime\prime} - 2 g^\prime g^{-1} g^\prime + \Tr\,
(g^{-1} g^\prime)\, g^\prime] + {\rm Ric} (g) - (d - 2)\,
g^\prime - \Tr \,(g^{-1} g^\prime)\, g & = & 0 \cr
\nabla_i\, \Tr \,(g^{-1} g^\prime) - \nabla^j g_{ij}^\prime  & = & 0 \cr
\Tr \,(g^{-1} g^{\prime\prime}) - \frac{1}{2} \Tr \,(g^{-1} g^\prime
g^{-1}
g^\prime) & = & 0 , \label{eqn}
\eea
where differentiation with respect to $\rho$ is denoted with a prime,
$\nabla_i$ is the covariant derivative constructed from the metric
$g$, and ${\rm Ric} (g)$ is the Ricci tensor of $g$.

These equations are solved order by order in $\r$. This is achieved
by differentiating the equations with respect to $\r$ and then setting 
$\r=0$. For even $d$, this process would have broken down at order $d/2$
if the logarithm was not introduced in (\ref{coord}). $h_{(d)}$
is traceless, $\Tr\, \gi h_{(d)}=0$,
and covariantly conserved, $\nabla^i h_{(d)ij}=0$.
We show in appendix C that $h_{(d)}$
is proportional to the metric variation of the 
corresponding conformal anomaly, i.e. it is proportional to the 
stress-energy tensor of the theory with action 
the conformal anomaly. In any dimension, only the trace of 
$g_{(d)}$ and its covariant divergence are determined.
Here is where extra data from the CFT are needed:
as we shall see, the undetermined part is specified 
by the expectation value of the dual stress-energy tensor.

We collect in appendix \ref{EinSol} the results for 
$g_{(n)}$, $h_{(d)}$ as well as the results for the trace and 
divergence $g_{(d)}$. In dimension $d$ the latter are 
the only constraints that equations (\ref{eqn}) 
yield for $g_{(d)}$. From this information
we can parametrize the indeterminacy by finding the most general
$g_{(d)}$ that has the determined trace and divergence. 

In $d=2$ and $d=4$ the equation for the coefficient 
$g_{(d)}$ has the form of a conservation law
\be \label{gA}
\nabla^ig_{(d)ij}=\nabla^iA_{(d)ij}~~,\qquad d=2, 4
\ee
where $A_{(d)ij}$ is a symmetric tensor explicitly constructed from
the coefficients $g_{(n)},~n<d$. 
The precise form of the tensor $A_{(d)ij}$ is given
in appendix A (eq.(\ref{Ad})). 
The integration of this equation obviously involves an
``integration constant'' $t_{ij}(x)$, a symmetric covariantly conserved
tensor the precise form of which can not be determined from 
Einstein's equations.

In two dimensions, we get \cite{SkSo} (see also \cite{bautier})
\be \label{g2}
g_{(2) ij}= \half (R\, g_{(0) ij} + t_{ij}),
\ee
where the symmetric tensor $t_{ij}$ should satisfy
\be \label{t2}
\nabla^i t_{ij}=0, \qquad \Tr\, t = - R.
\ee

In four dimensions we obtain\footnote{From now on we will suppress 
factors of $g_{(0)}$. For instance,
$\Tr\, g_{(2)} g_{(4)}= \Tr\, [g_{(0)}^{-1}g_{(2)}g_{(0)}^{-1} g_{(4)}]$. 
Unless we explicitly mention to the contrary, indices will be raised and
lowered with the metric $g_{(0)}$, all contractions will 
be made with this metric.}
\be \label{g4}
g_{(4)ij}={1 \over 8} g_{(0)ij} \,[(\Tr\, g_{(2)})^2-\Tr\, g_{(2)}^2] + 
\half (g_{(2)}^2)_{ij} - {1 \over 4} g_{(2)ij}\, \Tr\, g_{(2)}
+t_{ij},
\ee
The tensor $t_{ij}$ satisfies
\be \label{t4}
\nabla^i t_{ij}=0, \qquad \Tr\, t 
= -{1 \over 4} [(\Tr\, g_{(2)})^2 - \Tr\, g_{(2)}^2].
\ee 

In six dimensions  the equation determining the 
coefficient $g_{(6)}$ is more subtle than the one in (\ref{gA}).
It given by
\be \label{dg6}
\nabla^ig_{(6)ij}=\nabla^i A_{(6)ij}
+{1\over 6}\Tr (g_{(4)} \nabla_jg_{(2)})
\ee
where the tensor $A_{(6)ij}$ is given in (\ref{Ad}). It 
contains a part which is antisymmetric in the indices $i$ and $j$. 
Since $g_{(6)ij}$ is by definition a symmetric tensor
the integration of equation (\ref{dg6}) is not straightforward. 
Moreover, it is not obvious that
the last term in (\ref{dg6}) takes a form of divergence of some local tensor. 
Nevertheless, this is indeed the case 
as we now show. Let us define the tensor $S_{ij}$, 
\begin{eqnarray} 
\label{Sij}
&&S_{ij}=\nabla^2C_{ij}-2R^{k \ l}_{\ i \ j} C_{kl}
+4(g_{(2)}g_{(4)}-g_{(4)}g_{(2)})_{ij}
+{1\over 10}(\nabla_i\nabla_jB
-g_{(0)ij}\nabla^2 B) \nonumber \\
&&\hspace{1cm}
+{2\over 5}g_{(2)ij}B+g_{(0)ij}(-{2\over 3}\Tr \, g_{(2)}^3
-{4\over 15}(\Tr \,g_{(2)})^3+
{3\over 5}\Tr \, g_{(2)}\Tr \, g^2_{(2)})~~,
\end{eqnarray}
where 
$$
C_{ij}=(g_{(4)}-{1\over 2}g^2_{(2)}+{1\over 4}g_{(2)}\Tr \,g_{(2)})_{ij}+
{1\over 8}g_{(0)ij}B~~,~~
B=\Tr \, g^2_2-(\Tr \, g_2)^2~~.
$$
The tensor $S_{ij}$ is a local function of the Riemann tensor. Its 
divergence and trace read
\be \label{dtS}
\nabla^iS_{ij}=-4\Tr (g_{(4)} \nabla_j g_{(2)})~~,~~
\Tr S=-8\Tr(g_{(2)}g_{(4)})~~.
\ee
With the help of the tensor $S_{ij}$ the equation (\ref{dg6}) 
can be integrated in a way similar to the $d=2,4$ cases. One obtains
\be \label{g6}
g_{(6)ij}=A_{(6)ij}-{1\over 24}S_{ij} +t_{ij}~~.
\ee
Notice that tensor $S_{ij}$ contains an antisymmetric part which 
cancels the antisymmetric part of
the tensor $A_{(6)ij}$ 
so that $g_{(6)ij}$ and $t_{ij}$ are symmetric tensors, as they should.
The symmetric tensor $t_{ij}$ satisfies 
\be \label{t6}
\nabla^it_{ij}=0~~,~~\Tr\, t =-{1\over 3}[{1\over 8}(\Tr g_{(2)})^3
-{3\over 8}\Tr g_{(2)}\Tr g^2_{(2)}
+{1\over 2}\Tr g^3_{(2)}-\Tr g_{(2)} g_{(4)}]~~.
\ee

Notice that in all three cases, $d=2,4,6$, the trace of $t_{ij}$ is 
proportional to the holographic conformal anomaly.
As we will see in the next section, the symmetric tensors $t_{ij}$ 
are directly related to the expectation value of the boundary 
stress-energy tensor.

When $d$ is odd the only constraint on the coefficient $g_{(d)ij}(x)$ is 
that it is conserved and traceless
\begin{equation}
\nabla^ig_{(d)ij}=0~~, \qquad \Tr \, g_{(d)}=0~~.
\label{odd}
\ee
So that we may identify
\be
g_{(d)ij}=t_{ij}~~.
\ee

\section{The holographic stress-energy tensor}
\setcounter{equation}{0}

We have seen in the previous section that given a conformal
structure at infinity we can determine an asymptotic expansion
of the metric up to order $\r^{d/2}$. We will now show that this term 
is determined by the expectation value of the dual stress-energy tensor. 

According to the AdS/CFT prescription, the expectation value
of the boundary stress-energy tensor is determined by 
functionally differentiating the on-shell gravitational 
action with respect to the boundary metric. 
The on-shell gravitational action, however, diverges. 
To regulate the theory we restrict
the bulk integral to the region $\r\geq\e$
and we evaluate the boundary term at $\r=\e$.
The regulated action is given by
\bea \label{regaction}
S_{\sm{gr,reg}}&=&{1 \over 16 \p \GN}\left[\int_{\r\geq\e} 
\d^{d+1}x\, \sqrt{G} \,(R[G] + 2 \L) 
- \int_{\r=\e} \d^d x \sqrt{\c}\, 2 K\right]= \\
&=&{1 \over 16 \p \GN} \int \d^d x \left[ 
\int_\epsilon \d\rho\,{d\over\rho^{d/2+1}}\,\sqrt{\det g(x,\rho )} 
+ {1\over \rho^{d/2}}
(-2 d \sqrt{\det g(x,\rho )}
+4 \rho\partial_\rho \sqrt{\det g (x,\rho )})|_{\rho=\epsilon}\right] 
\nonumber
\eea
Evaluating (\ref{regaction}) for the solution we obtained in the 
previous section we find that the divergences 
appears as $1/\e^k$ poles plus a logarithmic divergence \cite{HS},
\be \label{regaction1}
S_{\sm{gr,reg}} = {l \over 16 \pi \GN} \int \d^d x \sqrt{\det g_{(0)}} \left( 
\epsilon^{-d/2} a_{(0)} + \epsilon^{-d/2+1} a_{(2)} + \ldots 
+ \epsilon^{-1} a_{(d - 2)} - \log \epsilon\, a_{(d)} \right) + \co(\e^0),
\ee
where the coefficients $a_{(n)}$ are local covariant expressions
of the metric $g_{(0)}$ and its curvature tensor. We give the
explicit expressions, up to the order we are interested in, in appendix B.

We now obtain the renormalized action by subtracting  
the divergent terms, and then removing the regulator,
\be \label{renaction}
S_{\sm{gr,ren}}[g_{(0)}]=\lim_{\e \to 0}{1 \over 16 \p \GN}
[S_{\sm{gr,reg}} -\int \d^d x \sqrt{\det g_{(0)}} \left( 
\epsilon^{-d/2} a_{(0)} + \epsilon^{-d/2+1} a_{(2)} + \ldots 
+ \epsilon^{-1} a_{(d - 2)} - \log \epsilon\, a_{(d)} \right)]
\ee
The expectation value of the stress-energy tensor of the 
dual theory is given by
\be \label{tij1}
\<T_{ij}\> = {2 \over \sqrt{\det \g0}} 
{\pa S_{\sm{gr,ren}} \over \pa g_{(0)}^{ij}}
=\lim_{\e \to 0} 
{2 \over \sqrt{\det g(x, \e)}} {\pa S_{\sm{gr,ren}} \over \pa g^{ij}(x,\e)}  
=\lim_{\e \to 0}\left( {1 \over \e^{d/2-1}}\, T_{ij}[\c]\right)
\ee
where $T_{ij}[\c]$ is the stress-energy tensor of the theory 
at $\r=\e$ described by the action in (\ref{renaction}) but before the
limit $\e \to 0$ is taken ($\c_{ij}=1/\e\, g_{ij}(x,\e)$ is the 
induced metric at $\r=\e$). 
Notice that the asymptotic expansion of the metric only 
allows for the determination of the divergences of the 
on-shell action. We can still obtain, however, a formula 
for $\<T_{ij}\>$ in terms of $g_{(n)}$
since, as (\ref{tij1}) shows, we only need to know
the first $\e^{d/2-1}$ orders in the expansion of $T_{ij}[\c]$.
 
The stress-energy tensor $T_{ij}[\c]$ contains two contributions,
\be \label{tij2}
T_{ij}[\c]=T^{\sm{\sm{reg}}}_{ij}+T^{\sm{\sm{ct}}}_{ij},
\ee
$T^{\sm{\sm{reg}}}_{ij}$ comes from the 
regulated action in (\ref{regaction}) and $T^{\sm{\sm{ct}}}_{ij}$ 
is due to the counterterms. The first contribution is equal to 
\be \label{regtij}
T_{ij}^{\sm{reg}}[\c]=-{1 \over 8 \p \GN} (K_{ij} - K \c_{ij})
=-{1 \over 8 \p \GN}\,(-\pa_\e g_{ij}(x,\e) + g_{ij}(x,\e)\, 
\Tr [g^{-1}(x,\e) \pa_\e g(x,\e)] + {1-d \over \e} g_{ij}(x,\e))
\ee
The contribution due to counterterms can be obtained from 
the results in appendix B. It is given by
\bea \label{counterT}
T^{\sm{ct}}_{ij}&=&-{1 \over 8 \p \GN} \left( (d-1) \c_{ij} + {1 \over (d-2)}
(R_{ij} - \half R \c_{ij}) \right.\nonu 
&&\left.-{1 \over (d-4) (d-2)^2}[-\nabla^2 R_{ij} + 2R_{ikjl} R^{kl} 
+{d-2 \over 2 (d-1)} \nabla_i \nabla_j R - {d \over 2 (d-1)} R R_{ij} \right.
\nonu
&&\left.- \half \c_{ij} (R_{kl} R^{kl} - {d \over 4 (d-1)} R^2 
- {1 \over d-1} \nabla^2 R)] - T^a_{ij} \log \e \right)
\eea
where $T^a_{ij}$ is the stress-energy tensor of the action 
$\int \d^d x\, \sqrt{\det \c}\, a_{(d)}$. As it is shown in Appendix C,
$T^{a}_{ij}$ is proportional to the tensor $h_{(d)ij}$ appearing
in the expansion (\ref{coord}).

The stress tensor $T_{ij}[g_{(0)}]$ is 
covariantly conserved with respect to the metric $g_{(0)ij}$. 
To see this, notice that each of $T^{\sm{reg}}_{ij}$
and $T_{ij}^{\sm{ct}}$ is separately covariantly conserved 
with respect to the induced metric $\c_{ij}$ at $\r=\e$:
for $T^{\sm{reg}}_{ij}$ one can check this by using the 
second equation in (\ref{eqn}), for $T_{ij}^{\sm{ct}}$
this follows from the fact that it was obtained by varying 
a local covariant counterterm. 
Since all divergences cancel in (\ref{tij1}),
we obtain that the finite part in (\ref{tij1}) 
is conserved with respect to the metric $g_{(0)ij}$.

We are now ready to calculate $T_{ij}$.
By construction (and we will verify this below) the 
divergent pieces cancel between $T^{\sm{reg}}$ and $T^{\sm{ct}}$.

\subsection{$d=2$}

In two dimensions we obtain
\be
\<T_{ij}\>={l \over 16 \p \GN}\, t_{ij}
\ee
where we have used (\ref{g2}) and (\ref{t2})
and the fact that $T^a_{ij}=0$ since $\int R$ is a topological 
invariant (and reinstated the factor of $l$).
As promised, $t_{ij}$ is directly related to the boundary stress-energy 
tensor. Taking the trace we obtain
\be
\<T^i_i\> = -{c \over 24 \p}\, R
\ee
where $c=3l/2\GN$, which is the correct conformal anomaly 
\cite{BrownHenneaux}. 

Using our results, one can immediately obtain the stress-energy tensor 
of the boundary theory associated with a given solution $G$
of the three dimensional Einstein equations: one needs
to write the metric 
in the coordinate system (\ref{coord}) and then use the formula
\be \label{T2}
\<T_{ij}\>={2 l \over 16 \p \GN}\, (g_{(2)ij} - g_{(0)ij}\,\Tr\, g_{(2)}).
\ee
{}From the gravitational point of view  this is the quasi-local stress 
energy tensor associated with the solution $G$.

\subsection{$d=4$}

To obtain $T_{ij}$ we first need to rewrite the expressions in 
$T^{\sm{ct}}$ in terms of $\g0$.
This can be done with the help of the relation
\be
R_{ij}[\c] = R_{ij}[\g0] + {1 \over 4}\, \e \left(2 R_{ik} R^k{}_{j}
-2 R_{ikjl} R^{kl} -{1 \over 3} \nabla_i \nabla_j R + \nabla^2 R_{ij}
-{1 \over 6} \nabla^2 R g_{(0) ij}\right) + \co(\e^2).
\ee

After some algebra one obtains,
\bea
\<T_{ij}[\g0]\>=-{1 \over 8 \p \GN} \lim_{\e \to 0}&&
\left[{1 \over \e} (-g_{(2)ij} + g_{(0)ij} \Tr\, g_{(2)}
+ \half R_{ij} - {1 \over 4} g_{(0)ij} R)
\right. \nonu
&&\left. +\log \e\, (-2 h_{(4)ij} - T^a_{ij}) \right. \nonu
&&\left.-2 g_{(4)ij} - h_{(4)ij} - g_{(2)ij} \Tr\, g_{(2)} - \half g_{(0)ij} 
\Tr\, g_{(2)}^2 \right. \nonu
&&\left.{1 \over 8}( R_{ik} R^k{}_{j}
-2 R_{ikjl} R^{kl} -{1 \over 3} \nabla_i \nabla_j R + \nabla^2 R_{ij}
-{1 \over 6} \nabla^2 R g_{(0) ij}) \right. \nonu
&&\left.-{1 \over 4} g_{(2)ij} R + {1 \over 8} g_{(0)ij}
(R_{kl} R^{kl} -{1 \over 6} R^2) \right].
\eea
Using the explicit expression for $g_{(2)}$ and $h_{(4)}$ given 
in (\ref{gexp}) and (\ref{h4}) one finds that both the 
$1/\e$ pole and the logarithmic divergence cancel.
Notice that had we not subtracted the logarithmic divergence from 
the action, the resulting stress-energy tensor would
have been singular in the limit $\e \to 0$. 

Using (\ref{g4}) and (\ref{t4}) and after some algebra we obtain
\be
\<T_{ij}\>=-{1 \over 8 \p \GN} [-2 t_{ij} -3 h_{(4)}].
\ee
Taking the trace we get
\be
\<T^i_i\>={1 \over 16 \p \GN} (-2 a_{(4)}),
\ee
which is the correct conformal anomaly \cite{HS}.

Notice that since $h_{(4)ij}=-\half T^a_{ij}$ the contribution 
in the boundary stress energy tensor proportional to $h_{(4)ij}$
is scheme dependent. Adding a local finite counterterm proportional
to the trace anomaly will change the coefficient of this term. 
One may remove this contribution from the boundary stress
energy tensor by a choice of scheme.

Finally, one can 
obtain the energy-momentum  tensor of the boundary theory
for a given solution $G$ of the five dimensional Einstein
equations with negative cosmological constant. It is given by 
\be \label{T4}
\<T_{ij}\>={4  \over 16 \p \GN} [g_{(4)ij}
-{1 \over 8} g_{(0)ij} [(\Tr\, g_{(2)})^2-\Tr\, g_{(2)}^2] -
\half (g_{(2)}^2)_{ij} + {1 \over 4} g_{(2)ij} \Tr\, g_{(2)}],
\ee
where we have omitted the scheme dependent $h_{(4)}$ terms. From 
the gravitational point of view  this is the quasi-local stress 
energy tensor associated with the solution $G$.

\subsection{$d=6$}

The calculation of the boundary stress tensor in $d=6$ case
goes along the same lines as in $d=2$ and $d=4$ cases although
it is technically involved.  Up to local traceless covariantly conserved term
(proportional to $h_{(6)}$) the results is 
\be \label{T6}
\<T_{ij}\>={3  \over 8 \p \GN}\, 
(g_{(6) ij}-A_{(6)ij}+{1\over 24}S_{ij})~~.
\ee
where $A_{(6)ij}$ is given in (\ref{Ad}) and $S_{ij}$ 
in (\ref{Sij}).
It is covariantly conserved and has the correct trace
\be \label{T6trace}
\<T^i_i\>={1 \over 8\p \GN}(-a_{(6)})~~,
\ee
reproducing correctly the conformal anomaly in six dimensions \cite{HS}.

Given an asymptotically AdS solution in six dimensions equation
(\ref{T6}) yields the quasi-local stress energy tensor associated
with it.

\subsection{$d=2k+1$}

In this case one can check that the counterterms only 
cancel infinities. Evaluating the finite part we get 
\be \label{Todd}
\<T_{ij}\>={d  \over 16 \p \GN}\, g_{(d)ij}.
\ee
where $g_{(d)ij}$ can be identified with a traceless 
covariantly conserved tensor $t_{ij}$.
In odd boundary dimensions there are no gravitational conformal
anomalies, and indeed (\ref{Todd}) is traceless. 
As in all previous cases, one can also read (\ref{Todd})
as giving the quasi-local stress energy tensor associated 
with a given solution of Einstein's equations.

\subsection{Conformally flat bulk metrics}

In this subsection we discuss a special case
where the bulk metric can be determined to all orders 
given only a boundary metric. It was shown in \cite{SkSo} that,
given a conformally flat boundary metric, 
equations (\ref{eqn}) can be integrated to all orders
if the bulk Weyl tensor vanishes\footnote{
In \cite{SkSo} it was proven that if the bulk metric satisfies 
Einstein's equations and it has a vanishing
Weyl tensor, then the corresponding  boundary
metric has to be conformally flat. The converse is not necessarily true: 
one can have Einstein metrics with non-vanishing Weyl tensor
which induce a conformally flat metric in the boundary.}.
We show that the extra condition in the bulk metric
singles out a specific vacuum of the CFT.

The solution obtained in \cite{SkSo} is given by
\be\label{KS}
g(x,\rho )=g_{(0)}(x)+g_{(2)}(x)\rho+g_{(4)}(x)\rho^2~~,~~g_{(4)}
={1\over 4}(g_{(2)})^2
\ee
where $g_{(2)}$ is given in (\ref{gexp}) (we consider $d>2$), 
and all other coefficients $g_{(n)}$, $n>4$ vanish.
Since $g_{(4)}$ and $g_{(6)}$ are now known, one can 
obtain a local formula for the dual stress energy tensor 
in terms of the curvature by using (\ref{g4}) and (\ref{g6}).

In $d=4$, using (\ref{g4}) and $g_{(4)}={1\over 4}(g_{(2)})^2$, one obtains
\be\label{CF1}
t_{ij}=t^{\sm{cf}}_{ij}\equiv 
-{1\over 4}(g_{(2)})^2_{ij}+{1\over 4}g_{(2)ij}\Tr \, g_{(2)}-
{1\over 8}g_{(0)ij}[(\Tr \, g_{(2)})^2-\Tr \, g^2_{(2)}]~~.
\ee
It is easy to check that trace of $t^{\sm{cf}}_{ij}$ reproduces (\ref{t4}).
Furthermore, by virtue of Bianchi's, one can show that $t^{\sm{cf}}_{ij}$
is covariantly conserved. It is well-known that the stress-energy tensor of a quantum field theory on a conformally flat spacetime is a local function of the curvature tensor (see for example \cite{Birrell-Davies}). Our equation (\ref{CF1}) reproduces the corresponding expression given in \cite{Birrell-Davies}.

In $d=6$, using (\ref{g6}) and $g_{(6)}=0$ we find
\bea \label{CF2}
&&t_{ij}=t^{\sm{cf}}_{ij}\equiv 
[ {1\over 4}g^3_{(2)}-{1\over 4}g^2_{(2)}\Tr \, g_{(2)}
+{1\over 8}g_{(2)}(\Tr g_{(2)})^2-{1\over 8}g_{(2)}\Tr \, g_{(2)} \nonu
&&\hspace{0.5cm}
+g_{(0)}({1\over 8}\Tr \, g_{(2)}\Tr \, g^2_{(2)}
-{1\over 12}\Tr \, g^3_{(2)}-{1\over 24}
(\Tr \, g_{(2)})^3)]_{ij}~~.
\eea
One can verify that the trace of $t^{\sm{cf}}_{ij}$ reproduces (\ref{t6}) 
(taking into account that $g_{(4)}={1\over 4} g^2_{(2)}$) 
and that $t^{\sm{cf}}_{ij}$ is covariantly conserved (by virtue of Bianchi's).

Following the analysis in the previous subsections
we obtain
\be
\<T_{ij}\>={d  \over 16 \p \GN}\, t^{\sm{cf}}_{ij}.
\ee
So, we explicitly see that the global condition we imposed on the bulk 
metric implies that we have picked a particular vacuum in the 
conformal field theory. 

Note that the tensors $t^{\sm{cf}}_{ij}$ in (\ref{CF1}), (\ref{CF2})
are local polynomial functions of the Ricci scalar and the 
Ricci tensor (but not of the Riemann tensor) of the metric $g_{(0)ij}$.
It is perhaps an expected but still a surprising result that in
conformally flat backgrounds the anomalous stress tensor 
is a local function of the curvature.

\section{Conformal transformation properties of the stress-energy tensor}
\setcounter{equation}{0}

In this section we discuss the conformal transformation 
properties of the stress-energy tensor. These can be 
obtained by noting \cite{ISTY} that conformal transformations
in the boundary originate from specific diffeomorphisms that 
preserve the form of the metric (\ref{coord}).
Under these diffeomorphisms $g_{ij}(x,\r)$ transforms infinitesimally 
as \cite{ISTY}
\be
\delta g_{ij}(x, \r) = 2 \s (1 - \r \pa_\r)\, g_{ij}(x, \r) +
\nabla_i a_j(x,\r) + \nabla_j a_i(x,\r),
\ee
where $a_j(x,\r)$ is obtained from the equation
\be
a^i(x,\r)=\half \int_0^\r \d\r' g^{ij}(x, \r') \pa_j \s(x).
\ee
This can be integrated perturbatively in $\r$,
\be
a^i(x,\r)=\sum_{k=1} a_{(k)}^i \r^k.
\ee
We will need the first two terms in this expansion,
\be \label{acoe}
a_{(1)}^i= \half \pa^i \s, \qquad
a_{(2)}^i=-{1 \over 4} g_{(2)}^{ij} \pa_j \s.
\ee

We can now obtain the way the $g_{(n)}$'s transform
under conformal transformations \cite{ISTY}
\bea \label{conftra}
&&\delta g_{(0)ij} = 2 \s g_{(0)ij}, \nonu
&&\delta g_{(2)ij} = \nabla_i a_{(1) j} + \nabla_j a_{(1) i} \nonu
&&\delta g_{(3)ij} = - \s g_{(3)ij}, \nonu
&&\delta g_{(4)ij} =-2 \s (g_{(4)} + h_{(4)}) 
+a^k_{(1)} \nabla_k g_{(2) ij}
+\nabla_i a_{(2) j}  + \nabla_j a_{(2) i}
+g_{(2) ik} \nabla_j a_{(1)}^k 
+g_{(2) jk} \nabla_i a_{(1)}^k \nonu
&&\delta g_{(5)ij} = - 3 \s g_{(3)ij},
\eea
where the term $h_{(4)}$ in $g_{(4)}$
is only present when $d=4$.
One can check from the explicit expressions for 
$g_{(2)}$ and $g_{(4)}$ in (\ref{gexp}) that 
they indeed transform as (\ref{conftra}).
An alternative way to derive the transformation rules
above is to start from (\ref{gexp}) and perform a
conformal variation.
In \cite{ISTY} the variations (\ref{conftra}) were
integrated leading to (\ref{gexp}) up to 
conformally invariant terms. 

Equipped with these results and the explicit form of the 
energy-momentum tensors, we can now easily calculate 
how the quantum stress-energy tensor transforms under conformal
transformations. We use the term ``quantum stress-energy 
tensor'' because it incorporates the conformal anomaly.
In the literature such transformation rules were obtained \cite{CC} by first 
integrating the conformal anomaly to an effective 
action. This effective action is a functional of the 
initial metric $g$ and of the conformal factor $\s$. It can be
shown that the difference between the stress-energy tensor
of the theory on the manifold with metric $g e^{2 \s}$ and the one
on the manifold with metric $g$ is given by  
the stress-energy tensor derived by varying the effective action 
with respect to $g$. 

In any dimension 
the stress-energy tensor transforms {\em classically} under 
conformal transformations as
\be \label{clasTr}
\delta \<T_{\m \n}\> = -(d-2)\, \s\, \<T_{\m \n}\>
\ee
This transformation law is modified by the quantum conformal
anomaly. In odd dimensions, where there is no
conformal anomaly, the classical transformation rule 
(\ref{clasTr}) holds also at the quantum level.
Indeed, for odd $d$, and by using (\ref{Todd}) and (\ref{conftra}),
one easily verifies that the holographic 
stress-energy tensor transforms correctly.
 
In even dimensions, the transformation (\ref{clasTr}) is modified.
In $d=2$, it is well-known that one gets an extra contribution
proportional to the central charge.
Indeed, using (\ref{T2}) and the formulae above we obtain
\be
\delta \<T_{ij}\>
={l \over 8 \p \GN}\, (\nabla_i \nabla_j \s - g_{(0)ij} \nabla^2 \s) 
={c \over 12}\, (\nabla_i \nabla_j \s - g_{(0)ij} \nabla^2 \s),
\ee
which is the correct transformation rule.

In $d=4$ we obtain,
\bea 
\delta \<T_{ij}\>&=&- 2 \s \<T_{ij}\> 
+{1 \over 4 \p \GN} \left(-2 \s h_{(4)}+
{1 \over 4} \nabla^k \s [\nabla_k R_{ij} 
- \half (\nabla_i R_{jk} + \nabla_j R_{ik}) - {1 \over 6} \nabla_k R g_{(0)ij}]
\right. \nonu 
&& \left.+{1 \over 48}(\nabla_i \s \nabla_j R + \nabla_i \s \nabla_j R) 
 +{1 \over 12} R (\nabla_i \nabla_j \s - g_{(0)ij} \nabla^2 \s) 
\right. \nonu 
&& \left.+{1 \over 8} [R_{ij} \nabla^2 \s 
- (R_{ik} \nabla^k \nabla_j \s + R_{jk} \nabla^k \nabla_i \s)
+ g_{(0)ij} R_{kl} \nabla^k \nabla^l \s] \right).
\eea
The only other result known to us is the result in \cite{CC},
where they computed the finite conformal transformation of the
stress-energy tensor but for a conformally flat metric $g_{(0)}$.
For conformally flat backgrounds, $h_{(4)}$ vanishes because 
it is the metric variation of a topological invariant. 
The terms proportional to a single derivative of $\s$  
vanish by virtue of Bianchi identities and the fact that the Weyl tensor 
vanishes for conformally flat metrics. 
The remaining terms, which only contain second derivatives 
of $\s$, can be shown to coincide with the
infinitesimal version of (4.23) in \cite{CC}.
 
One can obtain the conformal transformation
of the stress energy tensor in $d=6$ in a similar fashion
but we shall not present this result here.
 
\section{Matter} \label{matter}
\setcounter{equation}{0}

In the previous sections we examined how spacetime is 
reconstructed (to leading order) holographically out of CFT data. In this 
section we wish to examine how field theory describing 
matter on this spacetime is encoded in the CFT. 
We will discuss scalar fields but the techniques are readily applicable
to other kinds of matter. 

The method we will use is the same as in the case of 
pure gravity, i.e. we will start by specifying the 
sources that are turned on, find how far we can go 
with only this information and then input more CFT data.
We will find the same pattern: knowledge of the sources
allows only for determination of the divergent part of the 
action. The leading  finite part (which depends on global issues
and/or the signature of spacetime) is determined by the 
expectation value of the dual operator. We would like 
to stress that in the approach we follow, i.e.
regularize, subtract all infinities by adding counterterms and finally 
remove the regulator to obtain the renormalized action,
all normalizations of the physical correlation functions
are fixed and are consistent with Ward identities. 

Other papers that discuss similar
issues include \cite{AreVol,Nojod,NiTa,marika2}.

\subsection{Dirichlet boundary problem for scalar fields in a fixed
gravitational background}

In this section we consider scalars on a fixed
gravitational background. This is taken to be of the 
generic form (\ref{coord}). In most of the literature
the fixed metric was taken to be that of standard AdS,
but with not much more effort one can consider
the general case. 

The action for massive scalar is given by
\be \label{mataction}
S_{\tnnn{M}}=\half \int \d^{d+1}x \,\sqrt{G}\,
\left( G^{\m \n} \pa_\m \F \pa_\n \F
+ m^2 \F^2 \right)
\ee
where $G_{\m \n}$ has an expansion of the form
(\ref{coord}). 

We take the scalar field $\F$ to have an expansion 
of the form
\be \label{F}
\F(x,\r)=\r^{(d-\D)/2}\, \f(x,\r), \qquad 
\f(x,\r)=\f_{(0)} + \f_{(2)} \r + ... ~,
\ee
where $\D$ is the conformal dimension of the dual operator. 
We take the dimension $\D$ to be quantized as 
$\D={d \over 2} + k, k=0,1,..$. This is often the case
for operators of protected dimension. For the case of
scalars that correspond to operators 
of dimensions  ${d \over 2}-1 \leq \D < {d \over 2}$
we refer to \cite{KleWit}.
Inserting (\ref{F}) in the field equation, 
\be
(-\Box_G + m^2) \F =0,
\ee
where $\Box_G \F = {1 \over \sqrt{G}} \pa_\m (\sqrt{G} G^{\m \n} \pa_\n \F)$, 
we obtain that the mass $m^2$ 
and the conformal dimension $\D$ are related as $m^2=(\D-d)\D$,
and that $\f$ satisfies
\be \label{phieq}
[-(d-\D) \pa_\r \log g\, \f + 2 (2 \D -d -2) \pa_\r \f 
-\Box_g \f] + \r [-2 \pa_\r \log g\, \pa_\r \f - 4 \pa^2_\r \f]=0.
\ee
Given $\f_{(0)}$ one can determine recursively $\phi_{(n)}, n>0$.
This is achieved by differentiating (\ref{phieq}) and setting 
$\r$ equal to zero.
We give the result for the first couple of orders in appendix D.
This process breaks down 
at order $\D-d/2$ (provided this is an integer, which we assume throughout
this section)
because the coefficient of $\phi_{(2 \D -d)}$ (the 
field to be determined) becomes zero. This is exactly 
analogous to the situation encountered for even $d$
in the gravitational sector. Exactly the same way as there, we introduce 
at this order a logarithmic term, i.e. the expansion of $\F$  now reads,
\be \label{fexp}
\F = \r^{(d-\D)/2}\, (\f_{(0)} + \r \f_{(2)} + ...) 
+ \r^{\D/2}\, (\phi_{(2 \D -d)} + \log \r\, \psi_{(2 \D -d)} + ...).
\ee
The equation (\ref{phieq}) now determines all terms up to 
$\phi_{(2 \D -d -2)}$, the coefficient of the logarithmic term 
$\psi_{(2 \D -d)}$, but leaves undetermined
$\phi_{(2 \D -d)}$. This is analogous to the situation 
discussed in section 2 where the term $g_{(d)}$ was undetermined. 
It is well known \cite{BKL,BKLT,KleWit} that precisely at order 
$\r^{\D/2}$ one finds the expectation value of the dual
operator. We will review this argument below, and also 
derive the exact proportionality coefficient. Our 
result is in agreement with \cite{KleWit}.

We proceed to regularize and then renormalize the theory.
We regulate by integrating in the bulk from $\r \geq \e$,{}\footnote{
This regularization for scalar fields in a fixed AdS background
was considered in \cite{Mvis,FMMR}. In these papers the 
divergences were computed in momentum space, but no counterterms
were added to cancel them. Addition 
of boundary counterterms to cancel infinities for scalar
fields was considered in \cite{Gordon}, and more recently 
in \cite{KleWit}.}
\bea \label{matreg}
S_{\tnnn{M}\sm{,reg}}&=&\half \int_{\r \geq \e} \d^{d+1} x\, \sqrt{G}
\left( G^{\m \n} \pa_\m \F \pa_\n \F
+ m^2 \F^2 \right) \nonu
&=&-\int_{\r=\e} \d^d x\, 
\sqrt{g(x,\e)} \e^{-\D+d/2}\, [\half\, (d-\D) \f^2(x,\e)
+ \e\, \f(x,\e) \pa_\e \f(x,\e)] \\
&=& \int \d^d x\, \sqrt{g_{(0)}}\,
[\e^{-\D+d/2} a^{\tnnn{M}}_{(0)} + \e^{-\D+d/2+1} a^{\tnnn{M}}_{(2)}
+ ... + \e\,  a^{\tnnn{M}}_{(2 \D -d +2)} - \log \e\, a_{(2 \D -d)}] 
+ \co(\e^0) \nonumber
\eea
Clearly, with $\D-d/2$ a positive integer there are finite number 
of divergent terms.
The logarithmic divergence appears exactly  
when $\D=d/2+k, k=0,1,..$, 
in agreement with the analysis in \cite{PeSk}, 
and is directly related to the logarithmic term in (\ref{fexp}).
The first few of the power law divergences read
\be
a^{\tnnn{M}}_{(0)}=-\half (d-\D) \f_{(0)}^2, \qquad
a^{\tnnn{M}}_{(2)}=-{1 \over 4} \Tr\, g_{(2)}\, \f_{(0)}^2 + (d-\D +1)\, \f_{(0)} \f_{(2)}.
\ee
Given a field of specific dimension it is straightforward to 
compute all divergent terms.

We now proceed to obtain the renormalized action
by adding counterterms to cancel the infinities,
\be
S_{\tnnn{M}\sm{,ren}}=\lim_{\e \to 0} [S_{\tnnn{M}\sm{,reg}} - \int \d^d x\, \sqrt{g_{(0)}}\,
[\e^{-\D+d/2} a^{\tnnn{M}}_{(0)} + \e^{-\D+d/2+1} a^{\tnnn{M}}_{(2)}
+ ... + \e\, a^{\tnnn{M}}_{(2 \D -d +2)} - \log \e\,  a_{(2 \D -d)}] 
\ee
Exactly as in the case of pure gravity, and since the 
regulated theory lives at $\r=\e$, one needs to rewrite the
counterterms in terms of the field living at $\r=\e$, i.e. 
the induced metric $\c_{ij}(x, \e)$ and the field $\F(x,\e)$,
or equivalently $g_{ij}(x,\e)$ and $\f(x,\e)$.
This is straightforward but somewhat tedious:
one needs to invert the relation between $\f$ and $\f_{(0)}$ 
and between $g_{ij}$ and $g_{(0) ij}$ to sufficiently high order.
This then allows to express all $\f_{(n)}$, and therefore
all $a_{(n)}^{\tnnn{M}}$, in terms of $\f(x,\e)$ and $g_{ij}(x,\e)$ 
(the $\f_{(n)}$'s are determined in terms  of $\f_{(0)}$
and $g_{(0)}$ by solving (\ref{phieq}) iteratively).
Explicitly, the first two orders read,
\bea \label{finiteact}
S_{\tnnn{M}\sm{,ren}}&=&\lim_{\e \to 0} \left[\half \int_{\r \geq \e} \d^{d+1} x \sqrt{G}
\left( G^{\m \n} \pa_\m \F \pa_\n \F + m^2 \F^2 \right) \right. \\
&&\left.+ \int_{\r=\e}\, \sqrt{\c}\, [{(d-\D) \over 2} \F^2(x, \e) +
{1 \over 2(2 \D -d -2)}\, (\F(x,\e) \Box_\c \F(x,\e) 
+ {d-\D \over 2 (d-1)} R[\c] \F^2(x,\e)) + ...] \right] \nonumber 
\eea
The addition of the first counterterm was discussed in \cite{KleWit}.
The action (\ref{finiteact}) with only 
the counterterms written explicitly is finite 
for fields of $\D < d/2 +2$. As remarked above, it is
straightforward to  obtain all counterterms needed in order 
to make the action finite for any field of any mass.  
These counterterms contain also logarithmic subtractions
that lead to the conformal anomalies discussed in \cite{PeSk}.
For instance, if $\D=\half d +1$, the coefficient 
$[2 (2 \D -d -2)]^{-1}$ in (\ref{finiteact}) is replaced 
by $-{1 \over 4} \log \e$.
An alternative way to derive the counterterms is to demand that 
the expectation value $\<O\>$ is finite. This holds in the case of pure 
gravity too, i.e. the counterterms can also be derived by requiring
finiteness of $\<T_{\m \n}\>$ \cite{BK}.

The expectation value of the dual operator is given by
\be \label{oexp}
\< O (x) \> = - {1 \over \sqrt{\det g_{(0)}}}
{\delta S_{\tnnn{M}\sm{,ren}} \over \delta \f_{(0)}} =
- \lim_{\e \to 0} {1 \over \sqrt{\det g(x,\e)}}
{\delta S_{\tnnn{M}\sm{,ren}} \over \delta \f(x,\e)}. 
\ee
Exactly as in the case of pure gravity, the expectation 
value receives a contribution both from the regulated part 
and from the counterterms. We obtain,
\be \label{oexp1} 
\< O (x) \> = (2 \D - d)\, \f_{(2 \D -d)} 
+ F(\f_{(n)}, \psi_{(2 \D -d)}, g_{(m)}), \qquad n<2 \D-d
\ee
where we used that $\f_{(2 \D -d)}$ is linear in $\f_{(0)}$
(notice that the action (\ref{mataction}) does not include 
interactions). $F(\f_{(n)},\psi_{(2 \D -d)}, g_{(m)})$ 
is a local function of $\f_{(n)}$ with
$n<2\D{-}d$, $\psi_{(2 \D -d)}$ and $g_{(m)}$. 
These terms are related to 
contact terms in correlation functions of $O$ with itself and
with the stress-energy tensor. Its exact form is straightforward 
but somewhat tedious to obtain (just use (\ref{finiteact}) and
(\ref{oexp})). 

As we have promised, we have shown that the 
coefficient $\f_{(2 \D -d)}$ is related with the 
expectation value of the dual CFT operator. 
In the case that the background geometry is the standard 
Euclidean AdS one can readily 
obtain $\f_{(2 \D -d)}$ from the unique 
solution of the scalar field equation 
with given Dirichlet boundary conditions.
One finds that  $\f_{(2 \D -d)}$ is proportional to 
(an integral involving) $\f_{(0)}$. Therefore, $\f_{(2 \D -d)}$
carries information about the 2-point 
function. The factor $(\D - d/2)$ is crucial
in order for the 2-point function to be 
normalized correctly \cite{FMMR}. We refer to \cite{KleWit}
for a detailed discussion of this point.

We finish this section by calculating the conformal 
anomaly associated with the scalar fields and in the 
case the background is (locally) standard AdS (i.e. $g_{(n)}=0$, 
for $0<n<d$). Equation (\ref{phieq}) simplifies and can be 
easily solved. One gets 
\bea
\f_{(2n)}&=&{1 \over 2n(2 \D -d -2n)}\, \Box_0 \f_{(2n-2)}, \nonu
\psi_{(2\D-d)}&=&-{1 \over 2 (2 \D- d)}\, \Box_0 \f_{(2\D-d-2)}=
-{1 \over 2^{2k} \G(k) \G(k+1)}\, (\Box_0)^k \f_{(0)} \label{psian},
\eea
where $k=\D-{d \over 2}$ and $\Box_0$ is the Laplacian 
of $g_{(0)}$. The regularized action written in terms
of the fields at $\r=\e$ contains the following explicit 
logarithmic divergence,
\be
S_{\tnnn{M}\sm{,reg}}=-\int_{\r=\e} \d^dx \,\sqrt{\c}\, 
[\log \e\, 
(\D - {d \over 2})\, 
\f(x,\e)\, \psi_{(2\D-d)}(x,\e)+\cdots]\, ,
\ee
where the dots indicate power law divergent and finite terms,
$\psi_{(2\D-d)}(x,\e)$ is given by (\ref{psian}) with 
$g_{(0)}$ replaced by $\c$ and $\f_{(0)}$ by $\f(x,\e)$.
Using the same argument as in \cite{HS} we obtain 
the matter conformal anomaly,
\be
\ca_{\tnnn{M}}=\half \left({1 \over 2^{2k-2} (\G(k))^2} \right) \f_{(0)} 
(\Box_0)^k \f_{(0)}.
\ee
This agrees exactly with the anomaly calculated in \cite{PeSk}
(compare with formulae (10), (37) in \cite{PeSk}).

\subsection{Scalars coupled to gravity} \label{back}

In the previous section we ignored the back-reaction 
of the scalars to the bulk geometry. The purpose of
this section is to discuss this issue. The action 
is now the sum of (\ref{action}) and (\ref{mataction}),
\be \label{totaction}
S=S_{\sm{gr}}+S_{\tnnn{M}}.
\ee
The gravitational field equation in the presence of matter reads
\be
R_{\m \n} - \half (R + 2 \L) G_{\m \n}= - 8 \p \GN T_{\m \n}
\ee
In the coordinate system (\ref{coord}) and with the 
scalar field having the expansion in (\ref{fexp}),
these equations read
\bea \label{eqnmatter}
\rho \,[2 g^{\prime\prime}_{ij} - 2 (g^\prime g^{-1} g^\prime)_{ij} + \Tr\,
(g^{-1} g^\prime)\, g^\prime_{ij} \,] &+& R_{ij} (g) - (d - 2)\,
g^\prime_{ij} - \Tr \,(g^{-1} g^\prime)\, g_{ij} =
\\  \hspace{3cm}
&=&- 8 \p \GN\, \r^{d-\D-1}\left[{(\D-d) \D \over d-1}\, \f^2\, g_{ij} 
+ \r\, \pa_i \f \pa_j \f\right], \nonu  
\nabla_i \Tr \,(g^{-1} g^\prime) - \nabla^j g_{ij}^\prime &=&
-16 \p \GN\, \r^{d-\D-1} \left[{d-\D \over 2}\, \f \pa_i \f 
+ \r\, \pa_\r \f \pa_i \f\right], \nonu
\Tr \,(g^{-1} g^{\prime\prime}) - \frac{1}{2} \Tr\, (g^{-1} g^\prime
g^{-1} g^\prime) &=& - 16 \p \GN\, \r^{d-\D-2}
\left[{d(\D-d)(\D-d+1) \over 4 (d-1)}\, \f^2 \right.\nonu
&+&\left. (d-\D)\, \r\, \f \pa_\r \f 
+ \r^2\, (\pa_\r \f)^2\frac{}{}\right], \nonumber 
\eea

If $\D>d$, the right-hand side diverges near the boundary
whereas the left-hand side is finite. Operators with dimension
$\D>d$ are irrelevant operators. Correlation functions of these operators
have a very complicated singularity structure at coincident points.
As remarked in \cite{Wit}, one can avoid such problems by considering the 
sources to be infinitesimal and to have disjoint support, so that these 
operators are never at coincident points.  
Requiring that the equations in (\ref{eqnmatter}) are satisfied to leading 
order in $\r$ yields 
\be
\f_{(0)}^2=0, 
\ee
which is indeed the prescription advocated in \cite{Wit}. 
 
If $\D \leq d$, which means that we deal with marginal or relevant 
operators, one can perturbatively calculate the back-reaction of the 
scalars to the bulk metric. At which order the leading back-reaction
appears depends on the mass of the field. For fields that 
correspond to operators of dimension $\D=d-k$ the leading 
back-reaction appears at order $\r^k$, except when $k=0$
(marginal operators), where the leading 
back-reaction is at order $\r$. 

Let us see how conformal anomalies arise in this
context. The logarithmic divergences are coming from 
the regulated on-shell value of the bulk integral in 
(\ref{totaction}). The latter reads
\bea \label{totreg}
S_{\sm{reg}}(\mbox{bulk}) &=& \int_{\r \geq \e} \d \r\, \d^d x\, \sqrt{G}\, 
[{d \over 8 \p \GN} - {m^2 \over d-1}\, \Phi^2] \nonu
&=&\int_{\r \geq \e} \d \r\, \d^d x\, {1\over \r}\, \sqrt{g(x,\r)}\,
[{d \over 16 \p \GN}\, \r^{-d/2} - {m^2 \over 2(d-1)}\, \f^2(x,\r)\, \r^{-k}]
\eea
where $k=\D-d/2$. We see that gravitational conformal anomalies
are expected when $d$ is even and matter conformal anomalies
when $k$ is a positive integer, as it should.

In the presence of sources the expectation value 
of the boundary stress-energy tensor is not conserved but 
rather it satisfies a Ward identity that relates its covariant divergence
to the expectation value of the operators that couple to the 
sources. To see this consider the generating functional
\be
Z_{\tnnn{CFT}}[g_{(0)}, \phi_{(0)}]=\big< 
\exp \int \d^d x\, \sqrt{g_{(0)}}\,[\half\, g_{(0)}^{ij} T_{ij} - \f_{(0)} O] \big>.
\ee
Invariance under infinitesimal diffeomorphisms,
\be 
\delta g_{(0)ij} = \nabla_i \xi_j + \nabla_j \xi_i, 
\ee
yields the Ward identity,
\be \label{WI}
\nabla^j \< T_{ij} \> = \< O \>\, \pa_i \f_{(0)}.
\ee 
As we have remark before, $\<T_{ij}\>$ has a dual meaning\,\cite{BK},
both as the expectation value of the dual stress-energy tensor 
and as the quasi-local stress-energy tensor of Brown and York.
The Ward identity (\ref{WI}) has a natural explanation from
the latter point in view as well. According to \cite{BrownYork}
the quasi-local stress-energy tensor is not conserved in the 
presence of matter but it satisfies
\be \label{BY}
\nabla^j \<T_{ij}\> = - \t_{i\r}
\ee
where $\t_{i\r}$  expresses the flow of matter energy-momentum  through 
the boundary. Evidently, (\ref{WI}) is of the form (\ref{BY}).

Solving the coupled system of equations (\ref{eqnmatter}) and (\ref{phieq})
is straightforward but somewhat tedious. 
The details differ from case to case. 
For illustrative purposes we present a sample calculation:
we consider the case of two-dimensional massless scalar field
($d=\D=2, k=1$). 

The equations to be solved are (\ref{phieq}) and (\ref{eqnmatter})
with $d=\D=2$ and the
expansion of the metric and the scalar field are given 
by (\ref{coord}) and (\ref{fexp}) (again with $d=\D=2$), respectively.
Equation (\ref{phieq}) determines $\psi_{(2)}$,
\be \label{psi}
\psi_{(2)}=-{1 \over 4} \Box_0 \f_{(0)}.
\ee
Equations (\ref{eqnmatter}) determine $h_{(2)}$, the trace of the 
$g_{(2)}$ and provide a relation  
between the divergence of $g_{(2)}$ and $\f_{(2)}$,
\bea \label{d2ex}
&&h_{(2)}=- 4 \p \GN \left(\pa_i \f_{(0)}\pa_j \f_{(0)}
-\half\, g_{(0)ij}\, (\pa \f_{(0)})^2 \right), \nonu
&&\Tr\, g_{(2)} = \half\, R + 4 \p \GN\, (\pa \f_{(0)})^2, \nonu
&&\nabla^i g_{(2)ij}=\pa_i \Tr\, g_{(2)} + 16 \p \GN\, \f_{(2)} \pa_i \f_{(0)}.
\eea
Notice that $g_{(2)}$ and $\f_{(2)}$ are still undetermined
and are related to the expectation values of the dual operators
(\ref{tij1}) and (\ref{oexp1}), respectively.
Notice that $h_{(2)}$ is equal to the stress-energy tensor of a 
massless two-dimensional scalar. 

Going back to (\ref{totreg}), we see that the second term 
drops out (since $m^2=0$) and one can use the result already 
obtained in the gravitational sector,
\be \label{ga}
\ca={1 \over 16 \p \GN} (-2 a_{(2)})= {1 \over 16 \p \GN} (- 2 \Tr\, g_{(2)})
= - {1 \over 16 \p \GN} R + \half \f_{(0)} \Box_0 \f_{(0)} 
- {1\over 2}\,\nabla_i(\f_{(0)}\nabla^i \f_{(0)}),
\ee
which is the correct conformal anomaly \cite{HS,PeSk} (the last term can be 
removed by adding a covariant counterterm).

The renormalized boundary stress tensor reads
\begin{equation}
\<T_{ij}(x)\>={1\over 8\pi \GN}
\left(g_{(2) ij}+h_{(2) ij}-g_{(0) ij} \Tr\, g_{(2)} \right)(x)
\label{bT}
\end{equation}
Its trace gives correctly the conformal anomaly (\ref{ga}). 
On the other hand, taking the covariant derivative of (\ref{bT}) we get
\begin{eqnarray}
\nabla^j \<T_{ij}\>=\<O(x)\>\, \partial_i \phi_0(x)~~\nonumber \\
\<O(x)\>=2(\phi_2(x)+\psi_2(x)).
\label{Ward1}
\end{eqnarray}
in agreement with equations (\ref{WI}) and (\ref{oexp1}).

\section{Conclusions}
\setcounter{equation}{0}

Most of the discussions in the literature
on the AdS/CFT correspondence are concerned with obtaining
conformal field theory correlation functions using 
supergravity. In this paper we started investigating the 
converse question: how can one obtain information 
about the bulk theory from CFT correlation functions?
How does one decode the hologram? 

Answering these questions in all generality, but within the 
context of the AdS/CFT duality, entails developing 
a precise dictionary between bulk and boundary 
physics. A prescription for relating bulk/boundary 
observables is already available \cite{Gubs,Wit},
and one would expect that it would allow us to
reconstruct the bulk spacetime from the boundary CFT.
The prescription of \cite{Gubs,Wit}, however, relates infinite quantities.
One of the main results of this paper is the systematic
development of a renormalized version of this prescription.
Equipped with it, and with no other 
assumption (except that the CFT has an AdS dual),
we then proceeded to reconstruct the bulk 
spacetime metric and bulk scalar fields to the
first non-trivial order.

Our approach to the problem is to start from the boundary 
and try to build iteratively bulk solutions. Within 
this approach, the pattern we find is the following: \newline
$\bullet$ Sources in the CFT determine an asymptotic expansion
of the corresponding bulk field near the boundary to high enough order 
so  that {\em all infrared divergences} of the bulk on-shell
action can be computed. This then allows to obtain a 
renormalized on-shell action by adding boundary counterterms 
to cancel the infrared divergences. \newline
$\bullet$ Bulk solutions can be extended one order 
further by using the 1-point function of the corresponding dual CFT 
operator.

In the case the bulk field is the metric, our results show
that a conformal structure at infinity is not in general
sufficient in order to obtain a bulk metric. The first 
additional information one needs is the expectation 
value of the boundary stress energy tensor.

As a by-product, we have obtained ready-to-use formulae
for the Brown-York quasi-local stress-energy tensor
for arbitrary solution of Einstein's equations with 
negative cosmological constant up to six dimensions. The six-dimensional 
result is particularly interesting because,
via AdS/CFT, provides new information about the 
still mysterious $(2,0)$ theory. Furthermore, we 
have obtained the conformal transformation properties
of the  stress-energy tensors. These transformation 
rules incorporate the trace anomaly and provide 
a generalization to $d>2$ of the well-known Schwartzian 
derivative contribution in the conformal 
transformation rule of the stress-energy tensor in $d=2$.

Our discussion extends straightforwardly to the case 
of different matter. We expect that in all cases
obstructions in extending the solution to the 
deep interior region will be resolved by additional
CFT data (including data about non-local observables such as
Wilson loops, Wilson surfaces etc.). 
An interesting case to study in this framework
is point particles \cite{HaSk}. Reconstructing the 
trajectory of the bulk point particle 
out of CFT data will present a model of how 
holography works with time dependent
processes. Furthermore, following \cite{HorItz},
one could study the interplay between causality and
holography. Another extension is to study renormalization group 
flows using the present formalism. This amounts
to extending the discussion in section 5.2 by
adding a potential for the scalars. 
Another application of our results is in the context 
of Randall-Sundrum (RS) scenarios \cite{RS}. 
Incorporating such a scenario in string theory, 
in the case the bulk space is AdS,
may yield a connection with the AdS/CFT duality \cite{herman,Wrs}. 
As advocated in \cite{Wrs}, 
one may view the RS scenario as $4d$ gravity 
coupled to a cut-off CFT. The regulated theory in 
our discussion provides a dual description of a
cut-off CFT. In this context, the 
counterterms are re-interpreted as providing 
the action for the bulk modes localized 
in the brane \cite{KS,Gub,GKL}. We see, for instance,
that the counterterms in (\ref{finiteact}) can be 
re-interpreted as an action for a bulk scalar mode localized on the 
brane.
 
\section*{Note added} 
As this paper was being finalized,  \cite{BFRS} 
appeared with some overlap with the results of
section 2.

\section*{Acknowledgments} We would like to thank G. 't Hooft 
for reading the manuscript and his useful remarks.
This research is supported in part by NSF grants PHY94-07194
and PHY-9802484. KS  would like to thank ITP in UCSB 
for hospitality during initial stages of this work. 
SS would like to thank the Theory Division at CERN
for the hospitality extended to him while this work was in progress.
 
\appendix{Asymptotic solution of Einstein's equations} \label{EinSol}
\setcounter{equation}{0}

In this appendix we collect the results for the solution of the
equations (\ref{eqn}) up to the order we are interested in.

{}From the first equation in (\ref{eqn}) one determines 
the coefficients $g_{(n)}$, $n \neq d$, in terms of $g_{(0)}$.
For our purpose we only need $g_{(2)}$ and $g_{(4)}$.
There are given by
\bea \label{gexp}
g_{(2)}{}_{ij} & = & \frac{1}{d - 2} \left( R_{ij} - \frac{1}{2 (d - 1)} 
R\, \g0{}_{ij} \right) \cr
g_{(4)}{}_{ij} & = & \frac{1}{d - 4} \left( - \frac{1}{8 (d - 1)} D_i
D_j R + \frac{1}{4 (d - 2)} D_k D^k R_{ij} \right . \cr
& & - \frac{1}{8 (d - 1) (d - 2)} D_k D^k R \g0{}_{ij} - \frac{1}{2 (d - 2)}
R^{kl} R_{ikjl} \cr
& & + \frac{d - 4}{2 (d - 2)^2} R_i{}^k R_{kj} + \frac{1}{(d - 1)(d -
2)^2} R R_{ij} \cr
& & \left. + \frac{1}{4 (d - 2)^2} R^{kl} R_{kl} \g0{}_{ij} - \frac{3 d}{16
(d - 1)^2 (d - 2)^2} R^2 \g0{}_{ij} \right) .
\eea
The expressions for $g_{(n)}$ are singular when $n=d$. One can obtain the trace
and the divergence of $g_{(n)}$ for any $n$ from the last two equations
in (\ref{eqn}). Explicitly, 
\bea
&&\Tr\, g_{(4)} = {1 \over 4}\, \Tr\, g_{(2)}^2, \qquad 
\Tr\, g_{(6)}= {2 \over 3}\, 
\Tr\, g_{(2)} g_{(4)}
-{1 \over 6}\, \Tr\, g_{(2)}^3 \nonu
&&\Tr\, g_{(3)}=0, \qquad  \qquad \ \ \ \Tr\, g_{(5)}=0,
\eea
and 
\bea
&& \nabla^i g_{(2)ij} = \nabla^i A_{(2) ij}, \qquad 
\nabla^i g_{(3)ij} =0, \qquad
\nabla^i g_{(4) ij} = \nabla^i A_{(4) ij} \nonu
&&\nabla^i g_{(5)ij} =0, \qquad
\nabla^i g_{(6) ij}= \nabla^i A_{(6) ij}
+ {1 \over 6} \Tr\, (g_{(4)} \nabla_j g_{(2)})\, ,
\label{gd}
\eea
where 
\bea
A_{(2) ij}&=& g_{(0) ij} \Tr\, g_{(2)} \label{Ad} \\
A_{(4) ij} &=& - {1 \over 8}[\Tr\, g_{(2)}^2 - (\Tr\, g_{(2)})^2]\, g_{(0) ij} 
+ \half (g_{(2)}^2)_{ij} - {1 \over 4}\, g_{(2) ij}\, \Tr\, g_{(2)} \nonu
A_{(6) ij} &=& {1 \over 3} \left(
2(g_{(2)} g_{(4)})_{ij}+(g_{(4)} g_{(2)})_{ij}-(g_{(2)}^3)_{ij} 
+ {1 \over 8}\,[\Tr\, g_{(2)}^2 - (\Tr\, g_{(2)})^2]\, g_{(2) ij} \right. \nonu
&-&\left.\Tr\, g_{(2)}\,[g_{(4)ij} - \half (g_{(2)}^2)_{ij}] 
-[{1 \over 8} \Tr\, g_{(2)}^2 \Tr\, g_{(2)} - {1 \over 24} (\Tr\, g_{(2)})^3
-{1 \over 6} \Tr\, g_{(2)}^3
+{1\over 2} \Tr \, (g_{(2)}g_{(4)})]\,g_{(0) ij} \right)\, . \nonumber
\eea

For even $n=d$ the first equation in (\ref{eqn}) determines  
the coefficients $h_{(d)}$. They are given by
\bea
&&h_{(2)ij}=0  \label{h2} \\
&&h_{(4)ij}
={1\over 2}g^2_{(2)ij}-{1\over 8}g_{(0)ij}\Tr \, g^2_{(2)}+{1\over 8}
(\nabla^k\nabla_ig_{(2)jk}+\nabla^k\nabla_jg_{(2)ik}
-\nabla^2 g_{(2)ij}-\nabla_i\nabla_j
\Tr \, g_{(2)}) \label{h4} \\
&&\hspace{1cm}={1 \over 8} R_{ikjl} R^{kl} + {1 \over 48} \nabla_i \nabla_j R
-{1 \over 16} \nabla^2 R_{ij} -{1 \over 24} R R_{ij} 
+ ({1 \over 96} \nabla^2 R + {1 \over 96} R^2 -{1 \over 32} R_{kl}R^{kl})
g_{(0) ij}\,
\nonu
&&h_{(6)ij}={2\over 3}(g_{(4)}g_{(2)}+g_{(2)}g_{(4)})_{ij}
-{1\over 3}g^3_{(2)ij}-{1\over 6}g_{(4)ij}\Tr\, g_{(2)}
+{1\over 6}g_{(0)ij}(3\Tr g_{(6)}-3\Tr g_{(2)}g_{(4)}+\Tr g^3_{(2)}) 
\nonumber \\
&&-{1\over12}[-{1 \over 4} \nabla_i \nabla_j \Tr g_{(2)}^2 
-\nabla^k \nabla_i g_{(4)jk} -\nabla^k \nabla_j g_{(4)ik}
+ \nabla^2 g_{(4)ij} \nonu
&&+g_{(2)}^{kl} 
[\nabla_l \nabla_i g_{(2)jk} +\nabla_l \nabla_j g_{(2)ik} -\nabla_l
\nabla_k g_{(2)ij}] \nonu
&&+{1 \over 2} \nabla^k \Tr g_{(2)} 
(\nabla_i g_{(2)jk} +\nabla_j g_{(2)ik} - \nabla_k g_{(2)ij}) \nonu
&&+\half \nabla_i g_{(2)kl} \nabla_j g_{(2)}^{kl} 
+\nabla_k g_{(2)il} \nabla^l g_{(2)j}{}^{k}
-\nabla_k g_{(2)il} \nabla^k g_{(2)j}{}^{l}]. \label{h6}
\end{eqnarray}

\appendix{Divergences in terms of the induced metric}
\label{div-ind}
\setcounter{equation}{0}

In this appendix we rewrite the divergent terms of the 
regularized action in terms of the induced metric at $\r=\e$.
This is needed in order to derive the contribution of 
the counterterms to the stress energy tensor.

The coefficients $a_{(n)}$ of the divergent terms in the regulated
action (\ref{regaction1}) are given by 
\bea
&&a_{(0)}=2(1-d), \qquad \qquad \qquad \qquad \qquad 
a_{(2)}=b_{(2)}(d)\, \Tr\, g_{(2)}, \\
&&a_{(4)}=b_{(4)}(d)\,
[(\Tr\, g_{(2)})^2 - \Tr\, g_{(2)}^2], \qquad
a_{(6)}=\left( {1 \over 8}\,\Tr\, g_{(2)}^3
-{3 \over 8}\, \Tr\, g_{(2)} \Tr\, g_{(2)}^2
+{1 \over 2}\, \Tr\, g_{(2)}^3
- \Tr\, g_{(2)} g_{(4)} \right), \nonumber 
\eea  
where $a_{(6)}$ is only valid in six dimensions and the numerical 
coefficients in $a_{(2)}$ and $a_{(4)}$ are given by
\be
b_{(2)}(d\neq2)=-{(d-4)(d-1) \over d-2}, \ \
b_{(2)}(d=2)=1, \ \ 
b_{(4)}(d\neq4)={-d^2 + 9d -16 \over 4 (d-4)}, \ \
b_{(4)}(d=4)=\half.
\ee
Notice that the coefficients $a_{(n)}$ are proportional 
to the expression for the conformal anomaly (in terms of $g_{(n)}$)
in dimension $d=n$ \cite{HS}.

The counterterms can be rewritten in terms of the induced metric
by inverting the relation between $\c$ and $\g0$ perturbatively in $\e$. 
One finds
\bea \label{convert}
\sqrt{\g0}&=&
\e^{d/2} \left(1 - \half\, \e\, \Tr\, \gi g_{(2)}
+{1 \over 8}\, \e^2\, [(\Tr\, \gi g_{(2)})^2 + \Tr\, (\gi g_{(2)})^2] 
+ \co(\e^3) \right)\sqrt{\c} \nonu
\Tr \,g_{(2)} &=& 
{1 \over 2 (d-1)} {1 \over \e} \left(R[\c] + 
{1 \over d-2} (R_{ij}[\c] R^{ij}[\c] - {1 \over 2(d-1)} R^2[\c])
+\co(R[\c]^3) \right) \nonu
\Tr\, g_{(2)}^2 &=& {1 \over \e^2} {1 \over (d-2)^2} 
\left(R_{ij}[\c] R^{ij}[\c] + {-3d+4 \over 4(d-1)^2} R^2[\c]
+\co(R[\c]^3) \right)
\eea
The terms cubic in curvatures in (\ref{convert}) give vanishing 
contribution in (\ref{tij1}) up to six dimensions.

Putting everything together we obtain that the counterterms, 
rewritten in terms of the induced metric, are given by
\be \label{ct}
S^{\sm{ct}}=-{1 \over 16 \p \GN} \int_{\r=\e} 
\sqrt{\c}\left[2(1-d) + {1 \over d-2} R - {1 \over (d-4) (d-2)^2}
(R_{ij} R^{ij} - {d \over 4 (d-1)} R^2) - \log \e\, a_{(d)} + ...\right]
\ee
where all quantities are now in terms of the induced metric, including the 
one in the logarithmic divergence. These are exactly the counterterms 
in \cite{BK,EJM,KLS} except that these authors did not include the 
logarithmic divergence. Equation (\ref{ct}) should be understood as 
containing only divergent counterterms in each dimension. This means that 
in even dimension $d=2k$ one should include only the first $k$ counterterms 
and the logarithmic one. In odd $d=2k+1$, only the first 
$k+1$ counterterms should be included. The logarithmic counterterms
appear only for $d$ even. The counterterms in (\ref{ct})
render the renormalized action finite up to $d=6$. This covers 
all cases relevant for the AdS/CFT correspondence. It is straightforward
but tedious to compute the necessary counterterms for $d>6$. From 
(\ref{ct}) one straightforwardly obtains (\ref{counterT}).
 
\appendix{Relation between $h_{(d)}$ and the conformal anomaly $a_{(d)}$}
\label{h-a}
\setcounter{equation}{0}

We show in this appendix that the tensor $h_{(d)}$ appearing in expansion
of the metric in (\ref{coord}) when $d$ is even is a multiple
of the stress tensor derived from the action
$\int a_{(d)}$. ($a_{(d)}$ is, up to a constant, the holographic 
conformal anomaly).

This can be shown by deriving the stress-energy tensor of the 
regulated theory at $\r=\e$ in two ways and then comparing the
results. In the first derivation one starts from 
(\ref{regaction}) and obtains the regulated stress-energy tensor 
as in (\ref{regtij}). Expanding $T_{ij}^{\sm{reg}}[\c]$ in $\e$
(keeping $g_{(0)}$ fixed) we find that there is a logarithmic divergence,
\be
T_{ij}^{\sm{reg}}[\c;\log] = {1 \over 8 \p \GN} \log \e \
({3\over 2 }d-1) h_{(d) ij}.
\ee
On the other hand, one can derive $T_{ij}^{\sm{reg}}[\c]$ starting from 
(\ref{regaction1}). One has to first rewrite the terms in
(\ref{regaction1}) in terms of the induced metric. This is done 
in the previous appendix. Once $T_{ij}^{\sm{reg}}[\c]$ has been
derived, we expand in $\e$. We find the following logarithmic
divergence
\be
T_{ij}^{\sm{reg}}[\c;\log] = 
{1 \over 8 \p \GN} \log \e \left( (1-d)h_{(d)ij}-T_{ij}^a,
\right)
\ee
where $T_{ij}^a$ is the stress-energy tensor of the action 
$\int \d^d x\, \sqrt{\det g_{(0)}}\, a_{(d)}$.
If follows that
\be
h_{(d) ij}=-{2 \over d} T_{ij}^a
\ee
We have also explicitly verified this relation by brute-force computation
in $d=4$.

\appendix{Asymptotic solution of the scalar field equation}
\label{as-sc}
\setcounter{equation}{0}

We give here the first two orders of the solution of
the equation (\ref{phieq})
\bea \label{mattersol}
&&\f_{(2)}= {1 \over 2 (2 \D -d -2)} 
\left(\Box_0 \f_{(0)} + (d-\D) \f_{(0)} \Tr\, g_{(2)}\right), \nonu
&&\f_{(4)}= {1 \over 4 (2 \D -d -4)} 
\left(\Box_0 \f_{(2)} - 2\, \Tr\, g_{(2)} \f_{(2)}
-\half (d-\D) \,[\Tr\, g_{(2)}^2\, \f_{(0)} - 2 \Tr\, g_{(2)}\, \f_{(2)}] \right. \nonu
&& \hspace{3.5cm}\left. 
-{1 \over \sqrt{g_{(0)}}}\, \pa_\m (\sqrt{g_{(0)}}\, g_{(2)}^{\m \n} 
\pa_\n \f_{(0)})
+ \half \pa^i \Tr\, g_{(2)} \pa_j \f_{(0)} \right)\, ,
\eea
where in $\Box_0$ the covariant derivatives are with respect to $g_{(0)}$. 

If $2 \D -d -2k=0$ one needs to introduce a logarithmic term
in order for the equations to have a solution, as discussed 
in the main text. For instance, when $\D=\half d +1$, $\f_{(2)}$
is undetermined, but instead one obtains for the coefficient of the 
logarithmic term,
\be
\psi_{(2)}=-{1 \over 4} 
\left(\Box_0 \f_{(0)} + ({d \over 2}-1)\, \f_{(0)} \Tr\, g_{(2)}\right).
\ee

\end{document}